\documentclass[letterpaper,journal]{IEEEtran}
\usepackage{amsmath,amsfonts}
\usepackage{algorithmic}
\usepackage{algorithm}
\usepackage{array}
\usepackage[caption=false,font=normalsize,labelfont=sf,textfont=sf]{subfig}
\usepackage{textcomp}
\usepackage{stfloats}
\usepackage{flafter}
\usepackage{url}
\usepackage{verbatim}
\usepackage{graphicx}
\usepackage[percent]{overpic}
\usepackage{cite}
\usepackage{xcolor}
\usepackage{afterpage}
\usepackage{etoolbox}
\newtheorem{proposition}{Proposition}
\newtheorem{theorem}{Theorem}

\begin{document}

\title{\LARGE Cross-Room Passive WiFi Tracking: Joint Estimation of Target Trajectory and Virtual Transmitter Location}

\author{Heping Wang,
        Zhongqin Wang, \IEEEmembership{Member, IEEE},\\
        Henk Wymeersch, \IEEEmembership{Fellow, IEEE}, and
        J. Andrew Zhang, \IEEEmembership{Senior Member, IEEE}\vspace{-3em}


\IEEEcompsocitemizethanks{
\IEEEcompsocthanksitem Heping Wang, Zhongqin Wang, and J. Andrew Zhang (corresponding author) are with the School of Electrical, Mechanical and Biomedical Engineering, University of Technology Sydney, Sydney 2007, Australia. E-mail: heping.wang@student.uts.edu.au, \{zhongqin.wang, andrew.zhang\}@uts.edu.au. Henk Wymeersch is with the Department of Electrical Engineering, Chalmers University of Technology, Gothenburg, Sweden. E-mail: henkw@chalmers.se.}
}

\markboth{Journal of \LaTeX\ Class Files,~Vol.~14, No.~8, August~2021}%
{Shell \MakeLowercase{\textit{et al.}}: A Sample Article Using IEEEtran.cls for IEEE Journals}


\maketitle

\begin{abstract}
Indoor bistatic WiFi sensing must often operate across rooms, where intervening walls block the direct line-of-sight (LOS) path between the transmitter (Tx) and receiver (Rx). Recovering a human trajectory from channel-state-information (CSI)-derived delay, angle-of-arrival (AoA), and Doppler estimates is difficult because the Tx position is often unknown and NLOS propagation is inconsistent with direct-path geometry. This paper presents CoTrack, a self-calibrating passive WiFi tracking scheme that does not require the Tx position. CoTrack extracts one dominant human-induced response and represents its transmitter-side propagation by a virtual Tx, which it estimates jointly with the human trajectory. Because similar measurements can be explained by different virtual-Tx--trajectory pairs, CoTrack uses multi-start nonlinear least squares and accepts an initialization only when the best-fitting candidates concentrate around a common virtual-Tx position. It then applies online alternating optimization for continuous tracking. We analyze the local identifiability of the joint estimation problem and derive the corresponding Cram\'er--Rao bound. Across six LOS and cross-room NLOS experiments, CoTrack achieves a median trajectory error of \(1.14\) m (\(80\)th percentile: \(1.42\) m) and a median virtual-Tx discrepancy of \(0.47\) m (\(80\)th percentile: \(0.54\) m).
\end{abstract}


\section{Introduction}
\label{sec:introduction}
\looseness=-1
\IEEEPARstart{W}{iFi} sensing has progressed from research prototypes toward standardization, with IEEE 802.11bf defining sensing procedures for detection, localization, and recognition over WiFi networks \cite{du2025overview80211bf,wu2025isac}. Because channel state information (CSI) is available from commodity WiFi devices, passive human tracking can reuse existing wireless links without cameras, wearable tags, or dedicated radars \cite{ma2019wifiCSI,wang2023widfs}. In a whole-home deployment, one WiFi transmitter (Tx), such as an access point or laptop, may serve multiple receivers (Rxs) located in different rooms or sensing areas \cite{shen2024multiroom}. Each Rx can use its local CSI measurements to sense nearby human motion, allowing the same Tx to support several spatially separated sensing zones. Some of these Tx--Rx links may cross intervening walls that completely block the direct line-of-sight (LOS) path, resulting in cross-room NLOS propagation \cite{gu2025csipose}. Reliable tracking over such links is therefore important \mbox{for practical whole-home WiFi sensing.}

\looseness=-1
Related WiFi sensing studies have investigated target tracking \cite{qian2018widar2,xie2019mdtrack,wang2023widfs,wang2025pointclouds,wang2026wifi}, activity recognition~\cite{lee2025wi}, vital-sign monitoring~\cite{li2024spacebeat}, and environmental sensing \cite{wu2025isac}. To make CSI usable for these sensing tasks, a core issue is to suppress phase distortions caused by Tx-Rx clock asynchrony and antenna/radio-frequency (RF)-chain mismatch. Existing phase compensation methods \cite{wu2024sensing} usually rely on a reference path, typically the Tx-Rx LOS path when available, so the CSI-derived estimates are relative to this reference path rather than absolute geometric quantities. Early Doppler-based methods~\cite{qian2017widar} convert Doppler shifts into velocity constraints and fuse multiple links to update the target position over time, but require accurate Tx-Rx geometry to recover a two-dimensional trajectory. Recent works instead combine delay, angle-of-arrival (AoA), and Doppler for passive tracking with a single-antenna transmitter and a three-antenna receiver \cite{wang2026wifi}. These methods use delay and AoA to constrain the target's bistatic range and Rx-side direction, while Doppler describes how the propagation distance changes over time. WiDFS first isolates the dominant Doppler component and then estimates its AoA and reflection distance for real-time tracking; WiDFS2.0 represents multiple body reflections as Doppler--AoA--range point clouds and tracks them with an extended Kalman filter \cite{wang2023widfs,wang2025pointclouds}. WiDFS2.5 extracts delay, AoA, and Doppler from CSI power and tracks the target using a measured Tx position~\cite{wang2025csipower}. These studies have advanced passive WiFi tracking, but require a known or calibrated transmitter-side geometric reference to recover absolute trajectories.

In contrast, passive WiFi tracking without a known or calibrated transmitter-side geometric reference remains underexplored. The resulting problem involves recovering the Tx-side reference together with the human trajectory from the same CSI observations rather than using the reference as a calibration input. Related graph-based localization methods jointly estimate the position of a mobile device together with propagation-induced delay biases~\cite{venus2023graph,venus2024graph} or unknown virtual anchors~\cite{sun2024dcslam,leitinger2019belief} to mitigate obstructed LOS and multipath effects. These methods track a radio-equipped mobile device rather than a device-free person. Other passive systems estimate an unknown transmitter using a mobile receiver~\cite{xu2024radio} or jointly localize multiple asynchronous sniffers while tracking a transmitting target~\cite{suraweera2020environment}. Learning-based methods \cite{chen2023device,kato2025multi} map WiFi measurements to locations directly, but require environment-specific training and degrade when the Tx position or environment changes. The challenge is amplified in cross-room NLOS sensing, where the LOS reference path is blocked and the effective reference deviates from the physical Tx. CoTrack instead uses a single fixed commodity-WiFi receiver to jointly estimate the unknown Tx-side NLOS reference and the trajectory of a device-free target from relative delay, AoA, and Doppler.

Accurate NLOS tracking without a known Tx position therefore presents two key challenges:

\begingroup\clubpenalty=10000
\textit{1) Unpredictable NLOS propagation.} When the direct LOS path from the Tx to the Rx-side sensing area is blocked, the human-induced signal comprises attenuated components produced by penetration, reflection, diffraction, and scattering. These components follow different routes and superpose at the Rx; their relative amplitudes and phases also vary with the environment and the person's position. The dominant route may consequently change during tracking, and the extracted response may represent a mixture of several NLOS routes rather than one physical path. A conventional single-path bistatic model therefore cannot be applied directly.
\par\endgroup

\textit{2) Tx--trajectory coupling.} Commodity transmitters and receivers are unsynchronized, so their CSI phases contain random clock offsets. Removing these offsets requires a reference path within the same CSI measurement; consequently, the resulting delay and AoA estimates are relative to that path rather than absolute geometric quantities \cite{kotaru2015spotfi,zhang2020calwifi,zubow2021phase}. They can be converted into absolute measurements only if the reference-path delay and AoA are known. In practice, the Rx may not know the physical Tx position. Even when it does, an NLOS reference signal can follow a different route whose delay and AoA cannot be inferred from direct Tx--Rx geometry. Unknown delay and angular biases therefore remain coupled with the transmitter-side geometry and the human trajectory.

\looseness=-1
To address these challenges, we propose CoTrack, a self-calibrating passive WiFi tracking scheme for LOS and cross-room NLOS scenarios that does not require a known Tx position. CoTrack represents the unknown transmitter-side propagation reference by a virtual Tx; in LOS, this reference can coincide with the physical Tx, whereas in NLOS it represents the dominant propagation route. CoTrack jointly estimates the virtual-Tx position and human trajectory rather than relying on prior geometric calibration. With a single-antenna Tx and a three-antenna Rx, it extracts relative delay, relative AoA, and Doppler from the human-induced CSI response and fits these measurements to a bistatic geometric model. The resulting joint problem is nonconvex because different virtual-Tx positions and trajectories can explain the same measurements with similar errors. CoTrack therefore refines multiple candidate initializations by local nonlinear least squares and accepts a solution only after sufficient motion has made the best candidates spatially consistent. Subsequent coherent processing intervals (CPIs) provide additional target positions, enabling online alternating updates of both the trajectory and virtual-Tx estimate.

\begin{figure}[!t]
    \centering
    \includegraphics[width=\columnwidth]{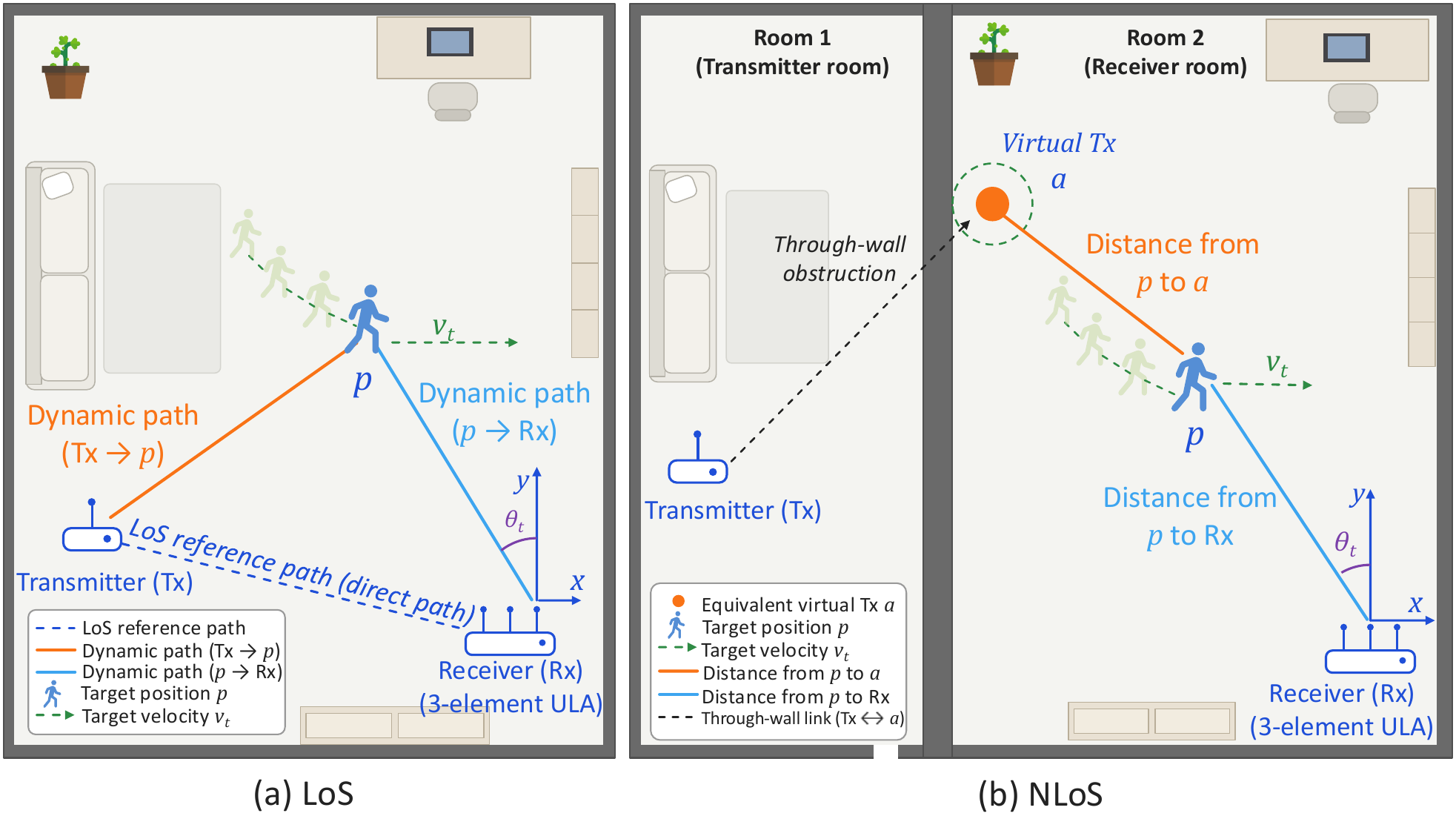}
    \vspace{-2.5em}
    \caption{System setups: (a) LOS deployment; (b) cross-room NLOS deployment, where the virtual Tx \(\mathbf a\) deviates from the physical Tx. The orange segment indicates the local ray of possible virtual-Tx \mbox{positions; \(\mathbf a\) is one candidate.}}
    \label{fig:system_setup}
    \vspace{-0.5em}
\end{figure}
The main contributions are as follows:

\begingroup\widowpenalty=10000\looseness=-1
\textit{1)} We formulate device-free passive WiFi tracking at a single fixed receiver with an unknown transmitter-side propagation reference. A virtual Tx approximates the Tx-to-person segment of the dominant NLOS response and is jointly estimated with the human trajectory. Unlike geometry-based passive WiFi trackers \cite{wang2023widfs,wang2025pointclouds} that assume known Tx and Rx positions, CoTrack uses relative delay, AoA, and Doppler to constrain the virtual Tx and trajectory as coupled unknowns. The same formulation applies to LOS and cross-room NLOS links.
\par\endgroup

\begingroup\widowpenalty=10000

\textit{2)} We analyze the local identifiability of the joint estimation problem. From the Fisher information matrix (FIM) and Cram\'er--Rao bound (CRB) for the virtual Tx and trajectory, we characterize how accuracy depends on motion extent, observation duration, and trajectory geometry. The analysis also identifies degenerate motion patterns that \mbox{make the joint problem ill-conditioned.}
\par\endgroup

\begingroup\looseness=-1
\textit{3)} We develop a multistage algorithm for the resulting nonconvex problem. CoTrack constructs multiple virtual-Tx initializations from the measured delay and AoA patterns, refines them by nonlinear least squares, and delays selection until the accumulated measurements yield a stable, spatially concentrated set of low-score candidates. It then updates the trajectory and virtual Tx through online alternating optimization.
\par\endgroup

\textit{4)} We evaluate CoTrack with a single-antenna transmitter, a three-antenna receiver, and \(20\)-MHz commodity WiFi CSI in LOS and cross-room NLOS scenarios. The experimental results show that CoTrack achieves a median trajectory error of \(1.14\) m (\(80\)th percentile: \(1.42\) m) and a median virtual-Tx discrepancy of \(0.47\) m (\(80\)th percentile: \(0.54\) m). Simulations validate identifiability and confidence-based initialization across motion spans and durations.

The rest of this paper is organized as follows. Section~\ref{sec:system_models} presents the system models, justifies the virtual Tx model, and overviews the proposed scheme. Section~\ref{sec:problem_formulation} formulates the joint estimation problem and analyzes its identifiability and CRB. Section~\ref{sec:cotrack} describes the CoTrack design, including virtual Tx initialization, confidence-based selection, online alternating updates, and complexity analysis. Section~\ref{sec:simulations} reports the simulation results, and Section~\ref{sec:experiments} reports the experimental evaluation. Section~\ref{sec:conclusion} concludes the paper, and the proofs and derivations are collected in the Appendix.

\section{System Models and Scheme Overview}
\label{sec:system_models}

\subsection{System Setups}
\label{sec:system_setup}

\begingroup\looseness=-1
As shown in Fig.~\ref{fig:system_setup}, a commodity WiFi Tx with one antenna transmits standard orthogonal frequency-division multiplexing (OFDM) packets, and an Rx with a three-element uniform linear array (ULA) extracts CSI. The Tx is uncontrolled: it neither shares a clock with the Rx nor reports its position. We define a two-dimensional coordinate system centered at the Rx, with the \(x\)-axis along the array and the \(y\)-axis along its broadside. Let \(\mathbf p\) and \(\mathbf a\) denote the target and virtual-Tx positions in this coordinate system. A walking person induces a dynamic Tx--target--Rx response whose delay, AoA, and Doppler support tracking. Its path length defines a bistatic ellipse whose foci are the Rx and the transmitter-side reference. In the LOS deployment of Fig.~\ref{fig:system_setup}(a), this reference coincides with the physical Tx. In the cross-room NLOS deployment of Fig.~\ref{fig:system_setup}(b), CoTrack represents the dominant human-induced response by a virtual Tx \(\mathbf a\), which need not coincide with the physical Tx. CoTrack estimates \(\mathbf a\) jointly with the human trajectory.
\par\endgroup

\subsection{NLOS Signal Models}
\label{sec:nlos_model}
Consider the cross-room NLOS Tx-Rx WiFi link in Fig.~\ref{fig:system_setup}(b), with a single transmit antenna and an \(M\)-antenna receiver. One OFDM packet yields one CSI snapshot over \(N_f\) subcarriers; \(N_p\) snapshots form a coherent processing interval (CPI). An estimation window comprises a sequence of consecutive CPIs whose measurements are used together to estimate the trajectory and virtual-Tx position. Let \(\mathbf C_n\in\mathbb C^{M\times N_f}\), \(n=1,\ldots,N_p\), denote the CSI matrix at the \(n\)-th snapshot, where \([\mathbf C_n]_{m,k}=\mathrm{CSI}_{n,k,m}\), \(m=1,\ldots,M\), and \(k=1,\ldots,N_f\). The raw CSI model is
\begin{equation}
    \mathbf C_n
    =
    \mathbf D_h
    \left(
    \mathbf H^{S}+\mathbf H_n^{X}
    \right)
    \mathbf D_n^{e},
    \label{eq:raw_csi_model}
\end{equation}
\begingroup\looseness=-1
where \(\mathbf H^{S}\) and \(\mathbf H_n^{X}\) are, respectively, the static and human-induced dynamic NLOS channel-frequency-response (CFR) matrices. The diagonal matrix \(\mathbf D_h\) models the antenna/RF-chain responses. The diagonal matrix \(\mathbf D_n^{e}\) models the timing offset (TO) and residual carrier-frequency offset (CFO); its \(k\)-th diagonal entry has unit magnitude and phase \(-2\pi f_k\tau_n^{\rm TO}+\psi_n^{\rm CFO}\). Here, \(\tau_n^{\rm TO}\) is the packet-wise TO, and \(\psi_n^{\rm CFO}\) is the phase accumulated by the residual CFO after coarse communication-oriented CFO estimation and compensation. Thus, \(\mathbf D_n^{e}\) is constant within snapshot \(n\) \mbox{but may vary across snapshots.}
\par\endgroup

Within this CPI, the dynamic CFR matrix can be written as
\begin{equation}
    \mathbf H_n^{X}
    =
    \sum_{\ell=1}^{L_X}
    \alpha_{\ell}^{X}
    e^{j2\pi f_{\ell}^{D}(n-1)T_s}
    \mathbf a_R(s_{\ell}^{X})
    \mathbf b_f^T(d_{\ell}^{X}),
    \label{eq:csi_dynamic_path}
\end{equation}
where \(L_X\) is the number of dynamic components, \(\alpha_{\ell}^{X}\) is the complex path gain, and \(T_s\) is the CSI snapshot interval. Let \(s_{\ell}^{X}=\sin\theta_{\ell}^{X}\) denote the spatial frequency. The vectors \(\mathbf a_R(s_{\ell}^{X})\) and \(\mathbf b_f(d_{\ell}^{X})\) are the receive-array and frequency steering vectors, with \(\big[\mathbf a_R(s)\big]_m=e^{-j2\pi(m-1)\Delta ds/\lambda}\) and \(\big[\mathbf b_f(d)\big]_k=e^{-j2\pi f_k d/c}\), where \(\Delta d\) is the antenna spacing, \(\lambda\) is the carrier wavelength, \(f_k\) is the \(k\)-th subcarrier frequency, and \(c\) is the propagation speed. The absolute path length \(d_{\ell}^{X}\) is not directly observable; person motion produces \(f_{\ell}^{D}=-\lambda^{-1}\frac{\mathrm d d_{\ell}^{X}}{\mathrm d t_{\rm c}}\), where \(t_{\rm c}\) denotes continuous time.

\subsection{Local Virtual-TX Approximation}
\label{sec:local_virtual_tx}

CoTrack tracks the dominant human-induced response in \(\mathbf H_n^{X}\). The following model applies to this response after TO removal and therefore omits TO. Let \(L(\mathbf p)\) be the physical propagation distance from the Tx to a person at \(\mathbf p\), measured along the NLOS route that produces this response. \mbox{The complete path length is}
\begin{equation}
d^{X}(\mathbf p)=L(\mathbf p)+\|\mathbf p\|_2,
\label{eq:physical_nlos_path}
\end{equation}

where the person-to-Rx segment is modeled as direct. This segment also determines the AoA at the Rx. The difficulty is that the transmitter-side function \(L(\mathbf p)\) is generally unknown and need not equal the Euclidean distance from the \mbox{physical Tx to the person.}

CoTrack locally approximates \(L(\mathbf p)\) by the distance from a virtual Tx \(\mathbf a\) to the person, plus a scalar path-length offset \(C\):
\begin{equation}
L(\mathbf p)\approx \|\mathbf p-\mathbf a\|_2+C.
\label{eq:virtual_tx_local_model}
\end{equation}
Here \(C\) compensates for the difference between the physical Tx-to-person path length and the virtual Tx-to-person straight-line distance. The following proposition shows that a ray of virtual-Tx positions provides the same first-order approximation around a reference target position \(\mathbf p_0\).
{
\begin{proposition}[Local virtual-Tx approximation]
\label{prop:existence}
Suppose that \(L(\mathbf p)\) is twice continuously differentiable near \(\mathbf p_0\) and satisfies the local unit-gradient condition \(\|\nabla L(\mathbf p)\|_2=1\). Define the local propagation direction
\begin{equation}
\mathbf u=\nabla L(\mathbf p_0).
\end{equation}
For any \(r>0\), place a virtual Tx a distance \(r\) behind \(\mathbf p_0\) along \(\mathbf u\), and choose the corresponding offset as
\begin{equation}
\mathbf a(r)=\mathbf p_0-r\mathbf u,
\qquad C(r)=L(\mathbf p_0)-r.
\label{eq:virtual_tx_ray}
\end{equation}
For each chosen \(r\), \(C(r)\) is fixed as \(\mathbf p\) varies near \(\mathbf p_0\). The physical and virtual models then have the same \mbox{value and gradient at \(\mathbf p_0\):}
\begin{equation}
\begin{aligned}
L(\mathbf p_0)
&=\|\mathbf p_0-\mathbf a(r)\|_2+C(r),\\
\nabla L(\mathbf p_0)
&=\nabla_{\mathbf p}\|\mathbf p-\mathbf a(r)\|_2\big|_{\mathbf p=\mathbf p_0}.
\end{aligned}
\label{eq:virtual_tx_first_order_match}
\end{equation}
Consequently, for \(\mathbf p\) near \(\mathbf p_0\),
\begin{equation}
L(\mathbf p)=\|\mathbf p-\mathbf a(r)\|_2+C(r)
+O(\|\mathbf p-\mathbf p_0\|_2^2).
\label{eq:virtual_tx_taylor_match}
\end{equation}
If, in addition, \(\nabla^2L(\mathbf p_0)=(\mathbf I-\mathbf u\mathbf u^T)/r\), the models share the Hessian at \(\mathbf p_0\) and differ by \mbox{\(o(\|\mathbf p-\mathbf p_0\|_2^2)\)}.
\end{proposition}
}

The proof is given in Appendix~\ref{app:virtual_tx_proof}. Equation~\eqref{eq:virtual_tx_ray} has a simple geometric interpretation: the local gradient \(\mathbf u\) specifies the direction from which the tracked NLOS route appears to arrive at the person, but a single local observation does not determine how far away the corresponding virtual source lies. Varying \(r\) therefore produces a ray \mbox{of locally equivalent virtual-Tx positions.}

The value equality in \eqref{eq:virtual_tx_first_order_match} makes the two transmitter-side path lengths identical at \(\mathbf p_0\), after including \(C(r)\). The gradient equality makes their path-length rates identical for any instantaneous target velocity and hence gives the same Doppler at \(\mathbf p_0\). Both models also give the same AoA because they use the same direct person-to-Rx segment. Away from \(\mathbf p_0\), however, different values of \(r\) generally predict different delay and Doppler evolution. Measurements collected over a sufficiently diverse trajectory can therefore distinguish the candidates.

\begingroup\looseness=-1
The physical meaning of the selected virtual Tx depends on the propagation mechanism. In LOS, choosing the physical Tx gives the exact transmitter-side distance. For a single planar specular reflection, the mirror image of the physical Tx is an exact virtual source. For penetration through a wall, a virtual Tx near the physical Tx together with a constant excess path length can provide a useful local approximation. With more complicated multipath, the estimated virtual Tx should be interpreted as an effective local geometric reference rather than as a physical source.
\par\endgroup

\vspace{-0.7em}
\subsection{Overview of the Proposed Scheme}
\label{sec:scheme_overview}

\begin{figure*}[!t]
    \centering
    \includegraphics[width=\textwidth]{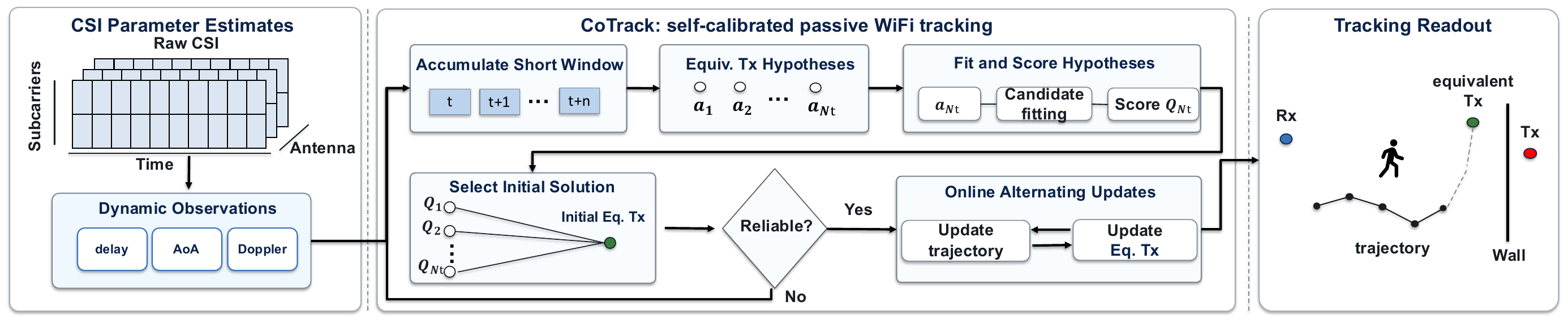}
    \vspace{-2.5em}
    \caption{Overview of CoTrack. A window of \(n+1\) CPIs generates \(N_t\) virtual-Tx hypotheses \(\mathbf a_i\), each with score \(Q_i\); \(N_t\) is independent of \(n\).}
    \label{fig:cotrack_framework}
    \vspace{-1.7em}
\end{figure*}

Fig.~\ref{fig:cotrack_framework} summarizes the CoTrack workflow. CoTrack first processes the raw CSI to remove packet-wise clock offsets and RF-chain phase offsets, and then estimates the delay, AoA, and Doppler of the human-induced response. Because neither the Tx position nor the delay/AoA references are known, CoTrack represents the transmitter-side geometry by a virtual Tx and estimates its position jointly with the human trajectory.

\begingroup\looseness=-1
To initialize this nonconvex problem, CoTrack generates \(N_t\) virtual-Tx hypotheses and fits each one to the delay, AoA, and Doppler estimates over a short observation window. Here, \(\mathbf a_i\) denotes the position of hypothesis \(i\), and \(Q_i\) is its score after refinement. CoTrack accepts an initial solution only when the low-score hypotheses are spatially concentrated. Otherwise, it accumulates more measurements and repeats the test. After initialization, CoTrack alternates between updating the trajectory and applying a conservative update to the virtual Tx and bias estimates. The same processing applies to LOS and NLOS deployments: NLOS-induced geometric displacement is absorbed into the virtual-Tx position, so no propagation-mode classifier or scenario-specific switch is required.
\par\endgroup

\section{Problem Formulation for Joint Tx-Position and Trajectory Estimation}
\label{sec:problem_formulation}

\begingroup\looseness=-1
This section relates CSI-derived relative delay, AoA, and Doppler to the unknown virtual Tx and human trajectory.
\par\endgroup

\vspace{-0.5em}
\subsection{Delay, AoA, and Doppler Estimation}

\subsubsection{Compensated Dynamic Signal}

Based on \eqref{eq:raw_csi_model}, the raw CSI must be processed to suppress \(\mathbf D_n^{e}\), compensate \(\mathbf D_h\), and remove \(\mathbf H^{S}\) before estimating the dynamic parameters in \(\mathbf H_n^{X}\).
Such phase compensation and parameter estimation can follow our previous works \cite{wang2023widfs,wang2025pointclouds,wang2026wifi} or other designs \cite{li2022csi,wu2024sensing}. In this work, we adopt a subspace-based compensation and super-resolution estimation scheme similar to \cite{zhao2025subspace}. It suppresses the clock-induced phases together with the static reference and estimates \mbox{the human-induced dynamic component, yielding}
\begin{equation}
    \mathbf R_n
    \approx
    \sum_{\ell=1}^{L_X}
    \beta_{\ell,n}
    e^{j2\pi f_{\ell}^{D}(n-1)T_s}
    \mathbf a_R(s_{\ell}^{\Delta})
    \mathbf b_f^T(d_{\ell}^{\Delta})
    +
    \mathbf N_n,
    \label{eq:relative_residual_model}
\end{equation}
where \(\beta_{\ell,n}\) is the residual path coefficient, \(\mathbf N_n\) collects noise, \(d_{\ell}^{\Delta}=d_{\ell}^{X}-d^{S}\) is the relative delay, \(s^{S}=\sin\theta^{S}\), \(s_{\ell}^{\Delta}=s_{\ell}^{X}-s^{S}=\sin\theta_{\ell}^{X}-\sin\theta^{S}\) is the relative spatial frequency. Here, \(d^{S}\) and \(\theta^{S}\) denote the path length and AoA of the static reference, respectively. Equation~\eqref{eq:relative_residual_model} represents each dynamic component by its relative delay \(d_{\ell}^{\Delta}\), relative spatial frequency \(s_{\ell}^{\Delta}\), and unchanged Doppler frequency \(f_{\ell}^{D}\).

\subsubsection{Biased Initial Sensing Parameter Estimates}
After phase compensation, the delay and AoA of each human-induced component remain referenced to an unknown static NLOS path and therefore contain unknown offsets. Whereas \(n\) indexes snapshots within a CPI, \(t\) indexes the successive CPI-level measurements used for tracking. For CPI \(t\), the parameter-estimation stage returns \((\hat d_{t,\ell}^\Delta,\hat s_{t,\ell}^\Delta,\hat f_{t,\ell}^D)\) for dynamic component \(\ell\). Let \(\ell\) denote the component selected for tracking, \mbox{and collect its measurements as}
\begin{equation}
\mathbf z_t
=
\begin{bmatrix}
    \hat d_{t,\ell}^{\Delta}\\
    \sin^{-1}\!\left(\hat s_{t,\ell}^{\Delta}\right)\\
    -\lambda\hat f_{t,\ell}^{D}
\end{bmatrix}
=
\begin{bmatrix}
    d_t\\
    \vartheta_t\\
    \nu_t
\end{bmatrix}
+
\begin{bmatrix}
    b_d\\
    b_\theta\\
    0
\end{bmatrix}
+
\begin{bmatrix}
    n_{d,t}\\
    n_{\theta,t}\\
    n_{\nu,t}
\end{bmatrix},
\label{eq:observation_model}
\end{equation}

\looseness=-1
where \(d_t\), \(\vartheta_t\), and \(\nu_t\) are the true delay-equivalent path length, AoA, and Doppler velocity, respectively. The second and third entries convert the spatial-frequency and Doppler estimates into the AoA and Doppler velocity forms used by the geometric model; the sign of the latter follows the phase convention in \eqref{eq:csi_dynamic_path}. Since the inverse sine accepts only values in \([-1,1]\), we clip \(\hat s_{t,\ell}^{\Delta}\) to this range before converting it to an angle. The constants \(b_d\) and \(b_\theta\) represent the unknown delay and angular offsets relative to the static NLOS reference \((d^S,\theta^S)\) used in phase compensation. The static-reference term \(-\sin\theta^S\) is constant in spatial frequency. Over a short CPI sequence, \(\vartheta_t\) varies only slightly, so the angle-domain offset \(\sin^{-1}(\sin\vartheta_t-\sin\theta^S)-\vartheta_t\) is approximately constant. We represent it by \(b_\theta\) and include the residual variation in \(n_{\theta,t}\). The terms \(n_{d,t}\), \(n_{\theta,t}\), and \(n_{\nu,t}\) are estimation errors. Since these biases arise from the unknown transmitter-side NLOS reference, the virtual Tx position, delay/AoA biases, and trajectory are coupled in the bistatic propagation geometry. Because no calibration trajectory is available initially, this coupling requires the virtual Tx and the initial trajectory to be estimated jointly.

\textit{Remark:} Under \eqref{eq:observation_model}, the likelihood is determined by the CPI-level estimates \(\{\mathbf z_t\}\) and their error covariance \(\boldsymbol\Sigma_e\). CoTrack can therefore use any CSI parameter-estimation method whose outputs satisfy this model.

Based on \eqref{eq:observation_model}, we next formulate the joint virtual-Tx and trajectory estimation problem. Section~\ref{sec:identifiability} then derives the CRB from the same initial-estimate model and its error covariance, without the motion regularizer or feasible-region bounds. We further quantify how errors in the CSI-derived parameters affect virtual-Tx and trajectory accuracy \mbox{through simulations in Section~\ref{sec:simulation_crb}.}

\vspace{-0.5em}

\subsection{Joint Optimization under Unknown Tx Geometry}
\label{sec:joint_optimization}

Let \(\mathbf p_t=[x_t,y_t]^T\) denote the human position at CPI \(t\) in the Rx-centered coordinates of Section~\ref{sec:system_setup}. For a sequence of \(T\) CPIs, let \(\mathcal W=\{t_1,\ldots,t_T\}\) be the CPI-index set used for joint estimation. A single virtual Tx models the dominant route throughout \(\mathcal W\), under the assumption that the route remains stable and the local approximation is valid. During initialization, the window grows until CoTrack finds a reliable solution; during online estimation, a fixed-length window slides forward. The corresponding measurement matrix \(\mathbf Z_{\mathcal W}\) and trajectory matrix \(\mathbf P_{\mathcal W}\) are
\begin{equation}
\begin{aligned}
    \mathbf Z_{\mathcal W}
    &=
    [\mathbf z_{t_1},\ldots,\mathbf z_{t_T}]^T
    \in\mathbb R^{T\times 3},\\
    \mathbf P_{\mathcal W}
    &=
    [\mathbf p_{t_1},\ldots,\mathbf p_{t_T}]^T
    \in\mathbb R^{T\times 2}.
\end{aligned}
\end{equation}
We formulate the estimation over \(\mathcal W\) as a regularized weighted least-squares problem that fits the delay, AoA, and Doppler estimates while penalizing implausible motion. Given the stacked estimates \(\mathbf Z_{\mathcal W}\), CoTrack jointly estimates the virtual Tx position \(\mathbf a=[x_a,y_a]^T\), the trajectory \(\mathbf P_{\mathcal W}\), and the bias vector \(\mathbf b=[b_d,b_\theta]^T\). The feasible set \(\Omega\) applies the target-position, bias, and virtual-Tx bounds in Table~\ref{tab:experimental_parameters}. The target bounds follow the approximate walking area and sensing range. The broad numerical bounds on \(\mathbf a\) allow the virtual Tx to \mbox{lie outside the physical room:}
\begin{equation}
\begin{aligned}
    \min_{(\mathbf a,\mathbf b,\mathbf P_{\mathcal W})\in\Omega}\quad
    &
    \left\|
    \mathbf E_{\mathcal W}(\mathbf Z_{\mathcal W};\mathbf a,\mathbf b,\mathbf P_{\mathcal W})
    \boldsymbol\Lambda
    \right\|_F^2
    +
    \lambda_r\mathcal R_r(\mathbf P_{\mathcal W}).
    \label{eq:joint_problem}
\end{aligned}
\end{equation}

Here, \(\mathbf E_{\mathcal W}\) stacks the differences between the measured and model-predicted delay, AoA, and Doppler values over the \(T\) CPIs. The diagonal matrix \(\boldsymbol\Lambda\) normalizes and weights the three residual types according to their nominal error scales and chosen reliabilities. The regularizer \(\mathcal R_r\) penalizes abrupt motion, and \(\lambda_r\) controls the tradeoff between measurement fit and trajectory smoothness. The \mbox{following subsections define these terms.}

\subsubsection{Prediction Model and Residual Matrix}
To obtain the velocity used in Doppler prediction and the acceleration used in motion regularization, let \(\mathbf D_1\) and \(\mathbf D_2\) be the central-difference matrices for a trajectory sampled every \(\Delta T\), where \(\Delta T\) is the interval between consecutive CPI-level estimates, distinct from the snapshot interval \(T_s\) in \eqref{eq:csi_dynamic_path}. For an interior row \(r\), \((\mathbf D_1\mathbf P_{\mathcal W})_r=(\mathbf p_{t_{r+1}}-\mathbf p_{t_{r-1}})^T/(2\Delta T)\) approximates the velocity, whereas \((\mathbf D_2\mathbf P_{\mathcal W})_r=(\mathbf p_{t_{r+1}}-2\mathbf p_{t_r}+\mathbf p_{t_{r-1}})^T/\Delta T^2\) approximates the acceleration. Thus, \(\mathbf V_{\mathcal W}=\mathbf D_1\mathbf P_{\mathcal W}\) and \(\mathbf A_{\mathcal W}=\mathbf D_2\mathbf P_{\mathcal W}\) are the velocity and acceleration stacks, respectively; \(\mathbf v_{t_r}^T\) is row \(r\) of \(\mathbf V_{\mathcal W}\), corresponding to CPI \(t_r\). Define \(\mathbf u_{R,t}=\mathbf p_t/\|\mathbf p_t\|_2\) and \(\mathbf u_{A,t}=(\mathbf p_t-\mathbf a)/\|\mathbf p_t-\mathbf a\|_2\) as the unit vectors from the Rx and virtual Tx toward the target, respectively. Differentiating the dynamic path length \(\|\mathbf p_t\|_2+\|\mathbf p_t-\mathbf a\|_2\) gives \((\mathbf u_{R,t}+\mathbf u_{A,t})^T\mathbf v_t\), which is the Doppler velocity prediction. The vector \(\mathbf h_t\) denotes the model-predicted delay, AoA, and Doppler values corresponding to \(\mathbf z_t\), including the unknown delay and angular biases,
\begin{equation}
    \mathbf h_t(\mathbf a,\mathbf b,\mathbf P_{\mathcal W})
    =
    \begin{bmatrix}
    \|\mathbf p_t\|_2+\|\mathbf p_t-\mathbf a\|_2-\|\mathbf a\|_2+b_d\\
    \operatorname{atan2}(x_t,y_t)+b_\theta\\
    (\mathbf u_{R,t}+\mathbf u_{A,t})^T\mathbf v_t
    \end{bmatrix}.
    \label{eq:predicted_observation}
\end{equation}
In the delay entry, \(b_d=C+\|\mathbf a\|_2-d^S\) accounts for the path-length offset \(C\) in the virtual-Tx approximation and the unknown static-reference path length \(d^S\). Equation~\eqref{eq:predicted_observation} shows why the unknown Tx geometry cannot be separated from tracking. The delay and Doppler entries depend on both \(\mathbf a\) and \(\mathbf p_t\), while the AoA entry depends on \(\mathbf p_t\) and the unknown angular reference. Therefore, different virtual Tx positions, bias values, and trajectories can yield similar residuals over a short motion window. For a given trial state, we abbreviate the prediction in \eqref{eq:predicted_observation} as \(\mathbf h_t\) and the residual in \eqref{eq:joint_problem} as \(\mathbf E_{\mathcal W}\). Stacking \(\mathbf h_t\) over \(\mathcal W\) gives the prediction matrix \(\mathbf H_{\mathcal W}\), and comparing it with the initial-estimate matrix \(\mathbf Z_{\mathcal W}\) \mbox{gives the residual matrix \(\mathbf E_{\mathcal W}\):}
\begin{equation}
    \begin{aligned}
    \mathbf H_{\mathcal W}
    &=
    [\mathbf h_{t_1},\ldots,\mathbf h_{t_T}]^T,\\
    \mathbf E_{\mathcal W}
    &=
    \mathbf Z_{\mathcal W}-\mathbf H_{\mathcal W},
    \end{aligned}
    \label{eq:stacked_residual}
\end{equation}
AoA residuals are wrapped to \((-\pi,\pi]\) to handle periodicity.
\subsubsection{Residual Weighting}
The weighting matrix in \eqref{eq:joint_problem} is
\begin{equation}
    \boldsymbol\Lambda
    =
    \operatorname{diag}
    \left(
    \frac{\sqrt{w_d}}{\sigma_d},
    \frac{\sqrt{w_\theta}}{\sigma_\theta},
    \frac{\sqrt{w_\nu}}{\sigma_\nu}
    \right).
\end{equation}

The residual scales \(\sigma_d,\sigma_\theta,\sigma_\nu\) normalize delay, AoA, and Doppler; the weights \(w_d,w_\theta,w_\nu\) set their relative \mbox{contributions to the fitting objective.}

\subsubsection{Motion Regularization}
The motion regularizer penalizes the second-order finite difference of the trajectory:
\begin{equation}
    \begin{aligned}
    \mathcal R_r(\mathbf P_{\mathcal W})
    &=
    \frac{1}{a_0^2}
    \|\mathbf A_{\mathcal W}\|_F^2
    =
    \frac{1}{a_0^2}
    \|\mathbf D_2\mathbf P_{\mathcal W}\|_F^2,
    \end{aligned}
    \label{eq:motion_regularizer}
\end{equation}
where \(a_0\) is the acceleration scale of the regularizer. Since \(\mathbf D_2\mathbf P_{\mathcal W}\) approximates the trajectory acceleration, \(\mathcal R_r\) penalizes rapid velocity changes. The weight \(\lambda_r\) balances trajectory smoothness against the CSI fit.

\vspace{-1em}
\subsection{Identifiability and CRB Analysis}
\label{sec:identifiability}

This subsection first uses the FIM and CRB to quantify virtual-Tx accuracy and then identifies the ambiguities and motion conditions that govern local identifiability.

\subsubsection{FIM and CRB}
{

\looseness=-1
Stack the unknowns as \(\boldsymbol\eta=[\mathbf a^T,\mathbf b^T,\mathbf p_{t_1}^T,\ldots,\mathbf p_{t_T}^T]^T\in\mathbb R^{4+2|\mathcal W|}\). Under the observation model in \eqref{eq:observation_model}, the Fisher information matrix is \(\mathbf F=\sum_{t\in\mathcal W}\mathbf J_t^{T}\boldsymbol\Sigma_e^{-1}\mathbf J_t\), where \(\mathbf J_t=\partial\mathbf h_t/\partial\boldsymbol\eta\); Appendix~\ref{app:fim_jacobians} gives the nonzero entries of \(\mathbf J_t\). After fixing the angular ambiguity by setting \(b_\theta=0\), the inverse on the nonsingular identifiable subspace gives the CRB. Interpreting the quadratic motion penalty as a Gaussian acceleration prior gives a local accuracy bound that includes this motion prior. The position and bias bounds constrain only the optimizer and are excluded from the FIM.

For the \(2\times2\) covariance submatrix \(\mathbf C_a\) of the virtual-Tx coordinates, \(B_a=\sqrt{\operatorname{tr}(\mathbf C_a)}\) is the Euclidean-position root-mean-square error (RMSE) bound. For this CRB, jointly scaling the delay, AoA, and Doppler error standard deviations by \(\kappa\) scales \(\mathbf F\) by \(\kappa^{-2}\) and \(B_a\) by \(\kappa\). Thus, \(B_a\) gives the \mbox{local virtual-Tx RMSE lower bound.}

\subsubsection{Local Identifiability and Global Ambiguity}

A common rotation of \(\mathbf a\) and the trajectory about the Rx leaves the predictions unchanged when absorbed by \(b_\theta\). For analysis, setting \(b_\theta=0\) fixes this rotational ambiguity. During estimation, the room and bias bounds constrain the orientation, and initialization determines which local solution is returned. Trajectory perturbations along the Rx line of sight can also be traded against \((\mathbf a,b_d)\) at first order, so feasibility bounds regularize the weak radial component of \(\mathbf a\).

\noindent\textit{Remark (Degenerate motion).} If the Tx-target direction remains constant over \(\mathcal W\), i.e., \(\mathbf u_{A,t}\equiv\mathbf u\), then shifting \(\mathbf a\) by \(\epsilon\mathbf u\) and adjusting \(b_d\) accordingly leaves all predictions unchanged. Thus, local virtual-Tx identifiability requires this direction to vary as the target moves. Appendix~\ref{app:degenerate_motion} gives the proof.

\looseness=-1
A singular FIM indicates directions to which the measurements are insensitive at first order. Separated minima with comparable losses represent a global ambiguity that local FIM analysis cannot resolve.

{
\begin{theorem}[Local sensitivity and candidate concentration]
\label{thm:curvature}
Set \(b_\theta=0\) and consider a solution strictly inside the feasible bounds. Let \(F_{\mathcal W}(\mathbf a)\) be the loss in \eqref{eq:joint_problem} after locally fitting \(b_d\) and the trajectory. The corresponding Gauss--Newton curvature matrix \(\mathbf H_a\) is defined in Appendix~\ref{app:profiled_curvature}.

(i) For a small change \(\delta\mathbf a\) in the virtual-Tx position, adjustments to \(b_d\) and the trajectory can cancel its effect on the residuals to first order if and only if \(\mathbf H_a\delta\mathbf a=\mathbf0\).

(ii) Near a local minimum with positive curvature in every direction, a small loss gap implies a small position difference. Specifically, suppose \(F_{\mathcal W}\) is twice continuously differentiable near an isolated local minimum \(\mathbf a^\star\), with \(\nabla^2F_{\mathcal W}(\mathbf a^\star)\succeq\lambda_a\mathbf I\), \(\lambda_a>0\). For sufficiently small \(\Delta_J>0\), every candidate \(\mathbf a_i\) in a sufficiently small neighborhood of \(\mathbf a^\star\) satisfying \(F_{\mathcal W}(\mathbf a_i)\le F_{\mathcal W}(\mathbf a^\star)+\Delta_J\) obeys \(\|\mathbf a_i-\mathbf a^\star\|_2\le\sqrt{2\Delta_J/\lambda_a}+o(\sqrt{\Delta_J})\).
\end{theorem}
}
The proof and CRB relation are given in Appendix~\ref{app:profiled_curvature}.
}

\section{Proposed CoTrack Scheme}
\label{sec:cotrack}

\looseness=-1
CoTrack solves \eqref{eq:joint_problem} using multiple starting points, local least-squares refinement, and a confidence check before online tracking. After initialization, alternating updates refine the trajectory and virtual-Tx estimate.

\subsection{Optimization Analysis and Solution Principle}
\subsubsection{Local Nonlinear Solver}
For a window \(\mathcal W\), define the state optimized in \eqref{eq:joint_problem} by stacking the virtual Tx \mbox{position, delay/AoA biases, and trajectory:}
\begin{equation}
    \mathbf x_{\mathcal W}
    =
    [\boldsymbol\chi^T,\operatorname{vec}(\mathbf P_{\mathcal W})^T]^T,
    \label{eq:vectorized_state}
\end{equation}
where \(\boldsymbol\chi=[x_a,y_a,b_d,b_\theta]^T\), and \(\operatorname{vec}(\cdot)\) stacks a matrix into a column vector. Using the residual and weighting definitions in Section~\ref{sec:joint_optimization}, \eqref{eq:joint_problem} can be written \mbox{as the nonlinear least-squares problem}
\begin{equation}
    \min_{\mathbf x_{\mathcal W}\in\Omega}
    \left\|
    \mathbf r_{\mathcal W}(\mathbf x_{\mathcal W})
    \right\|_2^2,
    \quad
    \mathbf r_{\mathcal W}(\mathbf x_{\mathcal W})
    =
    \begin{bmatrix}
    \operatorname{vec}(\mathbf E_{\mathcal W}\boldsymbol\Lambda)\\
    \sqrt{\lambda_r}\operatorname{vec}(\mathbf D_2\mathbf P_{\mathcal W})/a_0
    \end{bmatrix},
    \label{eq:vectorized_ls}
\end{equation}
where \(\Omega\) is the feasible set defined in \eqref{eq:joint_problem}. Given an initial state, CoTrack solves \eqref{eq:vectorized_ls} by damped Gauss--Newton updates
\begin{equation}
    \begin{aligned}
    \Delta\mathbf x^{(g)}
    &=
    \arg\min_{\Delta\mathbf x}
    \left\|
    \mathbf r^{(g)}
    +
    \mathbf J^{(g)}\Delta\mathbf x
    \right\|_2^2
    +
    \mu_{g}\|\Delta\mathbf x\|_2^2,\\
    \mathbf x^{(g+1)}
    &=
    \Pi_{\Omega}
    \left(\mathbf x^{(g)}+\Delta\mathbf x^{(g)}\right),
    \end{aligned}
    \label{eq:gn_local_step}
\end{equation}
where \(\mathbf x^{(g)}\) is the state for window \(\mathcal W\) at iteration \(g\), \(\mathbf r^{(g)}=\mathbf r_{\mathcal W}(\mathbf x^{(g)})\), \(\mathbf J^{(g)}=\partial\mathbf r_{\mathcal W}/\partial\mathbf x_{\mathcal W}|_{\mathbf x^{(g)}}\), \(\mu_{g}\) is the damping parameter, and \(\Pi_{\Omega}(\cdot)\) projects the update onto the feasible set. This update gives a local solution for a given initial state.

\subsubsection{Ambiguity and Candidate Concentration}
The ambiguities in Section~\ref{sec:identifiability} make initialization from a single starting point unreliable. CoTrack therefore refines multiple virtual-Tx candidates and starts online tracking when the low-score candidates are spatially concentrated and the best candidate has sufficient score contrast. The same steps apply in LOS and NLOS, since the virtual Tx represents the NLOS geometry.

\subsection{Initial Virtual-Tx Candidates}
Two sets of starting points are used: one around the Rx and one around a coarse center defined by the current delay/AoA estimates. Let \(t_0\) denote the first CPI with a detected human-induced dynamic component, and let \(\mathcal U_{\rm I}=\{u_1,u_2,\ldots\}\) denote the CPI indices at which initialization is evaluated as the window length increases. The interval \(\Delta_{\rm I}^{\rm ep}\) specifies the time between consecutive evaluations. At an evaluation time \(u\in\mathcal U_{\rm I}\), CoTrack uses the accumulated CPI window
\begin{equation}
    \mathcal W_{\rm I}(u)=\{t:t_0\le t\le u\}.
    \label{eq:initialization_cpi_window}
\end{equation}

For any angle \(\theta\), define the unit direction vector \(\mathbf u(\theta)\) as \([\sin\theta,\cos\theta]^T\). The current relative delay/AoA \mbox{estimates provide a coarse center}
\begin{equation}
    \mathbf c(u)
    =
    \operatorname{med}_{t\in\mathcal W_{\rm I}(u)}
    \hat d_{t,\ell}^{\Delta}\mathbf u\!\left(\sin^{-1}\!\left(\hat s_{t,\ell}^{\Delta}\right)\right).
    \label{eq:observation_center}
\end{equation}
Here \(\operatorname{med}\) denotes the componentwise median over the CPIs in \(\mathcal W_{\rm I}(u)\). The biased delay and AoA make \(\mathbf c(u)\) a search center for virtual-Tx starting points, not a target-position estimate. Let \(\widetilde{\mathcal R}_a\) and \(\widetilde{\Omega}_a\) contain the candidate offset radii \(\tilde r\) and Rx-frame bearings \(\tilde\alpha\) from \(\mathbf c(u)\):
\begin{equation}
    {
    \mathcal A_{\rm est}(u)
    =
    \{\mathbf c(u)+\tilde r\mathbf u(\tilde\alpha)
    \mid
    (\tilde r,\tilde\alpha)
    \in
    \widetilde{\mathcal R}_a\times\widetilde{\Omega}_a\}.
    }
    \label{eq:observation_candidate_set}
\end{equation}
For the Rx-centered starts, let \(\mathcal R_a\) and \(\Omega_a\) contain the candidate radii \(r\) and bearings \(\alpha\), respectively, measured from the \mbox{Rx. The global starts are}
\begin{equation}
    {
    \mathcal A_g
    =
    \{r\mathbf u(\alpha)
    \mid
    (r,\alpha)\in\mathcal R_a\times\Omega_a\}.
    }
    \label{eq:global_candidate_set}
\end{equation}
The multi-start initialization set is the union
\begin{equation}
    \mathcal A_0(u)=\mathcal A_g\cup\mathcal A_{\rm est}(u).
    \label{eq:anchor_candidate_set}
\end{equation}

Let \(N_t(u)=|\mathcal A_0(u)|\), and write \(\mathcal A_0(u)=\{\mathbf a_i^{(0)}(u)\}_{i=1}^{N_t(u)}\). Here \(i\) identifies the candidate, the superscript \((0)\) denotes its value before local refinement, and \(u\) \mbox{is the initialization evaluation time.}

\begin{figure*}[!t]
    \centering
    \includegraphics[width=0.495\textwidth]{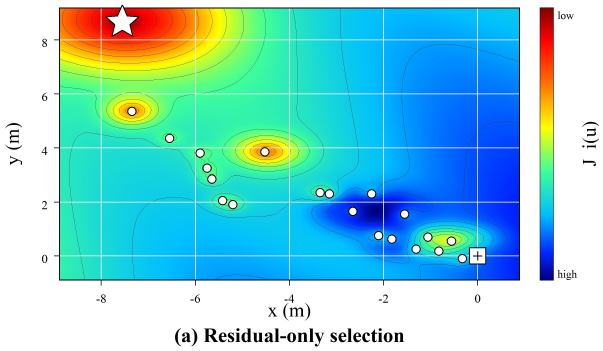}
    \hfill
    \includegraphics[width=0.495\textwidth]{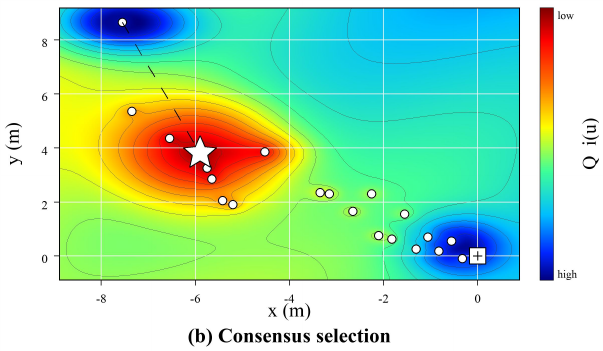}
    \vspace{-1.8em}
    \centerline{\footnotesize
    \(\circ\) virtual Tx candidate \qquad
    \(\star\) selected virtual Tx \qquad
    \raisebox{0.1ex}{\fbox{\scriptsize +}} receiver}
    \vspace{-0.4em}
    \caption{Virtual-Tx selection from the candidate score landscape. A residual-only rule may select an isolated low-loss point, while confidence-based acceptance also considers candidate spread and score contrast.}
    \label{fig:anchor_score_landscape}
    \vspace{-1.5em}
\end{figure*}

\subsection{Multi-Start Refinement}
\looseness=-1
For every virtual Tx start \(\mathbf a_i^{(0)}(u)\), CoTrack completes the state \eqref{eq:vectorized_state} with an initial trajectory and initial biases, and then refines it by solving \eqref{eq:vectorized_ls}.

The initial trajectory is shared by all virtual Tx starts. Let \(\bar d(u)\) denote the median relative-delay estimate over \(\mathcal W_{\rm I}(u)\), i.e.,
\(\bar d(u)=\operatorname{med}_{t\in\mathcal W_{\rm I}(u)}\hat d_{t,\ell}^{\Delta}\). For \(t\in\mathcal W_{\rm I}(u)\),
\begin{equation}
    \begin{aligned}
    \rho_t^{(0)}(u)
    &=
    \Pi_{[\rho_{\min},\rho_{\max}]}
    \left(\rho_0+\kappa_d(\hat d_{t,\ell}^{\Delta}-\bar d(u))\right),\\
    \mathbf p_t^{(0)}(u)
    &=
    \rho_t^{(0)}(u)\mathbf u\!\left(\sin^{-1}\!\left(\hat s_{t,\ell}^{\Delta}\right)\right),
    \end{aligned}
    \label{eq:trajectory_initialization}
\end{equation}
where \(\Pi_{[\rho_{\min},\rho_{\max}]}(\cdot)\) clips its argument to the initialized Rx-to-target distance bounds \([\rho_{\min},\rho_{\max}]\), \(\rho_0\) is the nominal Rx-to-target distance, and \(\kappa_d\) controls how the relative-delay variation changes the initial range. Stacking \(\mathbf p_t^{(0)}(u)\) over \(t\in\mathcal W_{\rm I}(u)\) gives \(\mathbf P^{(0)}(u)\). Equations~\eqref{eq:observation_candidate_set}--\eqref{eq:trajectory_initialization} specify starting points using the fixed settings in Table~\ref{tab:experimental_parameters}.

For each virtual Tx start \(\mathbf a_i^{(0)}(u)\), CoTrack then initializes the delay bias by median residual matching:
\begin{equation}
    \begin{aligned}
    b_{d,i}^{(0)}(u)
    &=
    \operatorname*{med}_{t\in\mathcal W_{\rm I}(u)}
    \Big[
    \hat d_{t,\ell}^{\Delta}
    -
    \|\mathbf p_t^{(0)}(u)\|_2
    \\
    &\qquad -
    \|\mathbf p_t^{(0)}(u)-\mathbf a_i^{(0)}(u)\|_2
    +
    \|\mathbf a_i^{(0)}(u)\|_2
    \Big],
    \\
    b_{\theta,i}^{(0)}(u)
    &=
    0.
    \end{aligned}
    \label{eq:bias_initialization}
\end{equation}
Since the initial trajectory lies along the measured directions, it carries no independent angular reference, so we initialize \(b_{\theta,i}^{(0)}(u)=0\) and refine it later.

The starting state \(\mathbf x_i^{(0)}(u)\) stacks \(\mathbf a_i^{(0)}(u)\), \(b_{d,i}^{(0)}(u)\), \(b_{\theta,i}^{(0)}(u)\), and \(\mathbf P^{(0)}(u)\) as in \eqref{eq:vectorized_state}.
CoTrack obtains a local solution \(\mathbf x_i^\star(u)\) by minimizing \(\|\mathbf r_{\mathcal W_{\rm I}(u)}(\mathbf x)\|_2^2\) over \(\Omega\) using \eqref{eq:gn_local_step}. Later evaluation times reuse the previous refined states as warm starts on the enlarged window, while newly generated candidates use the cold initialization above.
The refined solution is denoted by \(S_i(u)=\{\mathbf a_i(u),\mathbf b_i(u),\mathbf P_i(u),J_i(u)\}\), with normalized fitting loss \(J_i(u)=\|\mathbf r_{\mathcal W_{\rm I}(u)}(\mathbf x_i^\star(u))\|_2^2/|\mathcal W_{\rm I}(u)|\).

\begin{figure}[t]
    \centering
    \includegraphics[width=\columnwidth]{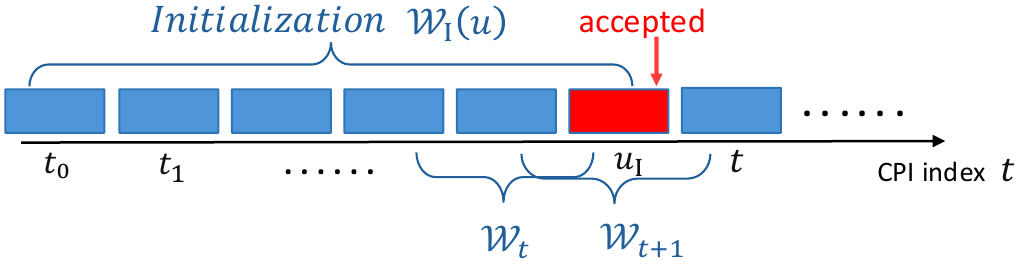}
    \vspace{-2em}
    \caption{\looseness=-1 Initialization and online sliding windows. After accepting \(u_{\rm I}\), CoTrack switches to sliding-window refinement.}
    \label{fig:window_timeline}
    \vspace{-1em}
\end{figure}

\subsection{Confidence-Based Initialization Decision}
\subsubsection{Candidate Scoring}
Given the refined set \(\{S_i(u)\}_{i=1}^{N_t(u)}\), the remaining task is to select a reliable initialization. A low fitting loss alone may not identify a reliable virtual Tx. CoTrack computes the following score for each candidate:
\begin{equation}
    Q_i(u)=J_i(u)+\lambda_p\Phi_p(\mathbf P_i(u)).
    \label{eq:candidate_score}
\end{equation}
Here \(\lambda_p\) weights the penalty for excessive speed and acceleration. The motion penalty is \(\Phi_p(\mathbf P_i(u))=N_{\rm fd}(u)^{-1}\sum_{r=1}^{N_{\rm fd}(u)}\bigl\{\max\!\bigl(\|(\mathbf D_1\mathbf P_i(u))_{r}\|_2/v_{\max}-1,0\bigr)^2+\allowbreak\max\!\bigl(\|(\mathbf D_2\mathbf P_i(u))_{r}\|_2/a_{\max}-1,0\bigr)^2\bigr\}\), where \(N_{\rm fd}(u)\) is the number of valid finite-difference rows, and \(v_{\max}\) and \(a_{\max}\) are the speed and acceleration limits. This soft penalty allows small motion-limit violations while balancing their magnitude against the CSI fit. Define \(Q_{\rm I}(u)=\min_i Q_i(u)\), and let \(S_{\rm I}(u)\) denote the refined solution that attains \(Q_{\rm I}(u)\).

\subsubsection{Initialization Confidence and Stopping Rule}
Motivated by the local concentration bound in Theorem~\ref{thm:curvature}(ii), CoTrack evaluates initialization confidence from the spatial spread and score contrast of low-score virtual-Tx candidates.

Concretely, let \(Q_{(1)}(u)\le\cdots\le Q_{(N_t(u))}(u)\) be the ordered scores, \(q_b=\lceil N_t(u)/4\rceil\), and
\(\mathcal B_Q(u)=\{i:Q_i(u)\le Q_{(q_b)}(u)\}\).
Let \(\bar{\mathbf a}_Q(u)\) denote the centroid of the candidate positions
\(\{\mathbf a_i(u):i\in\mathcal B_Q(u)\}\), and define their root-mean-square spread as
\(s_{\rm I}(u)=\big(|\mathcal B_Q(u)|^{-1}
\sum_{i\in\mathcal B_Q(u)}\|\mathbf a_i(u)-\bar{\mathbf a}_Q(u)\|_2^2\big)^{1/2}\).
The score contrast \(g_Q(u)=[Q_{(q_b)}(u)-Q_{(1)}(u)]/[Q_{(q_b)}(u)+\epsilon_Q]\), with \(\epsilon_Q>0\), gives
\begin{equation}
    C_b(u)
    =
    g_Q(u)
    \left(1+\frac{s_{\rm I}(u)}{r_c}\right)^{-1},
    \label{eq:initialization_confidence}
\end{equation}
where \(r_c\) normalizes distance and \(g_Q(u)\in[0,1]\) compares the best score \(Q_{(1)}(u)\) with \(Q_{(q_b)}(u)\). If \(C_b(u)\) is low, CoTrack continues to enlarge the window. Fig.~\ref{fig:anchor_score_landscape} contrasts the two cases. In Fig.~\ref{fig:anchor_score_landscape}(a), the minimum-score candidate is isolated while the other low-score candidates are dispersed, producing a large \(s_{\rm I}(u)\); thus, a low fitting loss alone is unreliable. In Fig.~\ref{fig:anchor_score_landscape}(b), the low-score candidates cluster around the selected virtual Tx and the best score is well below \(Q_{(q_b)}(u)\), yielding smaller \(s_{\rm I}(u)\), stronger \(g_Q(u)\), and higher \(C_b(u)\).

\looseness=-1
CoTrack accepts the first evaluation time passing both tests:
\begin{equation}
    u_{\rm I}=\min\{u\in\mathcal U_{\rm I}\setminus\{u_1\}:\Delta_{\rm I}(u)<\epsilon_{\rm I},\;
    C_b(u)\ge C_{\min}\},
    \label{eq:initialization_stop}
\end{equation}
Here, \(\Delta_{\rm I}(u)=|Q_{\rm I}(u)-Q_{\rm I}(u^{-})|/[Q_{\rm I}(u^{-})+\epsilon_Q]\), \(u^{-}\) is the preceding evaluation time, and \(\epsilon_Q\) prevents division by zero. The parameters \(\epsilon_{\rm I}\) and \(C_{\min}\) are the score-stability tolerance and confidence threshold, respectively. Thus, \(S_{\rm I}(u_{\rm I})\) initializes online tracking only when the best score is stable and the confidence score meets the threshold.

\begin{algorithm}[t]
\caption{CoTrack self-calibration and online tracking}
\label{alg:cotrack}
\begin{algorithmic}[1]
\REQUIRE CSI estimate stream \(\{\mathbf z_t\}\)
\ENSURE Virtual Tx, delay/AoA biases, and trajectory
\STATE Set \(t_0\) to the first CPI with a human-induced component.
\FOR{each initialization evaluation time \(u\in\mathcal U_{\rm I}\)}
    \STATE Form \(\mathcal W_{\rm I}(u)\) by \eqref{eq:initialization_cpi_window} and \(\mathcal A_0(u)\) by \eqref{eq:anchor_candidate_set}.
    \FOR{each \(\mathbf a_i^{(0)}(u)\in\mathcal A_0(u)\)}
        \STATE Initialize or warm-start \(\mathbf x_i^{(0)}(u)\).
        \STATE Refine to \(\mathbf x_i^\star(u)\) by \eqref{eq:gn_local_step} and compute \(Q_i(u)\).
    \ENDFOR
    \STATE Select \(S_{\rm I}(u)\) with minimum \(Q_i(u)\); compute \(C_b(u)\).
    \IF{\eqref{eq:initialization_stop} is satisfied}
        \STATE Accept \(u_{\rm I}=u\); initialize from \(S_{\rm I}(u_{\rm I})\); break.
    \ENDIF
\ENDFOR
\FOR{each online CPI \(t>u_{\rm I}\)}
    \STATE Update \(\mathcal W_t\) and warm-start from the previous state.
    \STATE Run \(I_o\) alternating updates by \eqref{eq:online_trajectory_update} and \eqref{eq:online_calibration_update}.
    \STATE Output the online state \(\mathbf x_t\).
\ENDFOR
\end{algorithmic}
\end{algorithm}

\vspace{-0.5em}
\subsection{Online Alternating Updates}
Let \(S_{\rm I}(u_{\rm I})=\{\mathbf a_{\rm I},\mathbf b_{\rm I},\mathbf P_{\rm I},J_{\rm I}\}\) be the accepted initialization and define
\(\boldsymbol\chi_{\rm I}=[\mathbf a_{\rm I}^T,\mathbf b_{\rm I}^T]^T\). After \(u_{\rm I}\) is accepted, CoTrack starts online alternating updates from the latest \(L_{\rm on}\)-CPI window ending at \(u_{\rm I}\), where \(L_{\rm on}\) is the maximum online window length. For each online CPI \(t\ge u_{\rm I}\), this sliding window is
\(\mathcal W_t=\{\max\{t_0,t-L_{\rm on}+1\},\ldots,t\}\). Each new CPI shifts \(\mathcal W_t\) forward by one CPI, as illustrated in Fig.~\ref{fig:window_timeline}. Let
\(\mathbf P_t=[\mathbf p_{\tau}]_{\tau\in\mathcal W_t}^T\) be the trajectory stack in this online window, and denote the corresponding residual by
\(\mathbf r_t(\boldsymbol\chi,\mathbf P_t)
=\mathbf r_{\mathcal W_t}([\boldsymbol\chi^T,\operatorname{vec}(\mathbf P_t)^T]^T)\).
For compact notation, define the online window loss as \(\mathcal L_t(\boldsymbol\chi,\mathbf P_t)=\|\mathbf r_t(\boldsymbol\chi,\mathbf P_t)\|_2^2\).

The first online state is initialized from \(S_{\rm I}(u_{\rm I})\), with \(\mathbf P_{\rm I}\) restricted to the CPIs in \(\mathcal W_{u_{\rm I}}\). For \(t>u_{\rm I}\), the update is warm-started from the previous online solution on the overlapping CPIs. At alternating iteration \(q\), CoTrack first refines the trajectory with the current virtual Tx and bias:
\begin{equation}
    \mathbf P_t^{(q+1)}
    =
    \arg\min_{\mathbf P_t}
    \mathcal L_t(\boldsymbol\chi_t^{(q)},\mathbf P_t).
    \label{eq:online_trajectory_update}
\end{equation}
Rather than replacing the current virtual-Tx and bias estimates, CoTrack moves only part of the way toward the new proposal:
\begin{equation}
    \boldsymbol\chi_t^{(q+1)}
    =
    (1-\gamma_\chi)\boldsymbol\chi_t^{(q)}
    +
    \gamma_\chi
    \operatorname*{arg\,min}_{\boldsymbol\chi}
    \mathcal L_t(\boldsymbol\chi,\mathbf P_t^{(q+1)}).
    \label{eq:online_calibration_update}
\end{equation}

where \(\gamma_\chi\) is the update step size. The virtual Tx is assumed constant within each online window, although its estimate is refined as the window advances. With \(\gamma_\chi=0.1\), each update combines \(10\%\) of the new proposal with \(90\%\) of the previous estimate, limiting sensitivity to short-window noise. Both least-squares subproblems use the bounded iteration in \eqref{eq:gn_local_step}. After \(I_o\) alternating iterations, the online state is \(\mathbf x_t=[(\boldsymbol\chi_t^{(I_o)})^T,\operatorname{vec}(\mathbf P_t^{(I_o)})^T]^T\). The last row of \(\mathbf P_t^{(I_o)}\) is the position estimate for CPI \(t\), and \(\boldsymbol\chi_t^{(I_o)}\) contains the \mbox{updated virtual-Tx position and biases.}

\begin{table}[!t]
    \caption{CoTrack parameter settings.}
    \label{tab:experimental_parameters}
    \centering
    \scriptsize
    \setlength{\tabcolsep}{1.5pt}
    \renewcommand{\arraystretch}{1.02}
    \begin{tabular}{@{}>{\raggedright\arraybackslash}p{0.34\columnwidth}>{\raggedright\arraybackslash}p{0.20\columnwidth}>{\raggedright\arraybackslash}p{0.38\columnwidth}@{}}
        \hline
        Parameter & Meaning & Value \\
        \hline
        Observation interval \(\Delta T\) & CPI spacing & \(0.05\) s \\
        \(\sigma_d,\sigma_\theta,\sigma_\nu\) & Residual scales & \(1.25\) m, \(28^\circ\), \(1.0\) m/s \\
        \(w_d,w_\theta,w_\nu\) & Fit weights & \(0.20\), \(0.35\), \(1.00\) \\
        \(\mathcal R_a\) & Global radius & \(\{2.3,5.8,8.9\}\) m \\
        \(\Omega_a\) & Global bearings & \(\{-112,-67,58,103\}^{\circ}\) \\
        \(\widetilde{\mathcal R}_a\) & Local radius & \(\{3.1,7.3\}\) m \\
        \(\widetilde{\Omega}_a\) & Local offsets & \(\{-138,-49,37,126\}^{\circ}\) \\
        \(\rho_0,\kappa_d,[\rho_{\min},\rho_{\max}]\) & Center range & \(3.5\) m, \(0.35\), \([0.5,7.5]\) m \\
        Target range & Target bounds & \([0.3,9.0]\) m \\
        Motion region & Walking area & \(x\in[-5,5]\) m, \(y\in[0,8]\) m \\
        Virtual-Tx coordinates & Numerical bounds & \(x_a,y_a\in[-60,60]\) m \\
        \(a_0,v_{\max},a_{\max}\) & Motion limits & \(6.0\) m/s\(^2\), \(2.75\) m/s, \(6.0\) m/s\(^2\) \\
        \(r_c,C_{\min}\) & Basin scale/test & \(0.75\) m, \(0.14\) \\
        \(b_d,b_\theta\) bounds & Bias bounds & \([-4,4]\) m, \([-120^\circ,120^\circ]\) \\
        \(\lambda_r,\lambda_p\) & Penalty weights & \(0.25\), \(0.45\) \\
        \(\Delta_{\rm I}^{\rm ep}\) & Initialization check interval & \(1.0\) s \\
        \(\epsilon_{\rm I},\epsilon_Q\) & Stop limits & \(0.03\), \(10^{-6}\) \\
        \(L_{\rm on},\gamma_\chi,I_o\) & Online update & \(60\) CPIs, \(0.10\), \(3\) \\
        \hline
    \end{tabular}
    \vspace{-1em}
\end{table}

\vspace{-1.0em}
\subsection{Computational Complexity}
Let \(C_{\rm GN}(T)\) denote the cost of one damped Gauss--Newton update over a \(T\)-CPI window. During initialization, CoTrack evaluates times \(u\in\mathcal U_{\rm I}\) up to the accepted \(u_{\rm I}\). At each time \(u\), it refines \(|\mathcal A_0(u)|\) candidates over \(\mathcal W_{\rm I}(u)\), with mean iteration count \(I_{\rm I}(u)\), for a total cost of
\begin{equation}
    \mathcal O
    \left(
	    \sum_{\substack{u\in\mathcal U_{\rm I}\\u\le u_{\rm I}}}
    |\mathcal A_0(u)|\,
    I_{\rm I}(u)\,
    C_{\rm GN}(|\mathcal W_{\rm I}(u)|)
    \right).
    \label{eq:initialization_complexity}
\end{equation}
The first evaluation is solved from cold starts, while later evaluations use warm-started states from the previous one, which reduces \(I_{\rm I}(u)\). Candidate scoring and basin-confidence evaluation require one pass over the refined trajectories and a sort over \(\mathcal A_0(u)\).
After initialization, each online CPI performs \(I_o\) alternating updates over \(\mathcal W_t\), with \(|\mathcal W_t|\le L_{\rm on}\). Let \(C_{\rm on}(T)\) denote the cost of one alternating update over a \(T\)-CPI window. The online per-CPI cost is
\begin{equation}
    \mathcal O\left(I_oC_{\rm on}(L_{\rm on})\right).
    \label{eq:online_complexity}
\end{equation}
With the parameters in Table~\ref{tab:experimental_parameters}, optimizer-only timings on an Apple M1 CPU are \(0.77\) s per warm-start initialization check and \(0.041\) s per online CPI (\(L_{\rm on}=60\), \(I_o=3\)).

\begin{figure*}[!t]
    \centering
    \newcommand{\crbcell}[2]{%
    \begin{minipage}[t]{0.326\textwidth}
        \centering
        \includegraphics[width=\linewidth]{#1}\\[-1.4ex]
        {\footnotesize #2}
    \end{minipage}}
        \crbcell{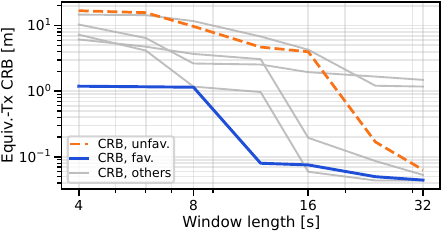}{(a) Window length}\hfill
        \crbcell{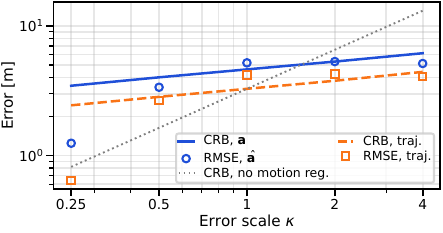}{(b) Parameter-estimation error scale}\hfill
        \crbcell{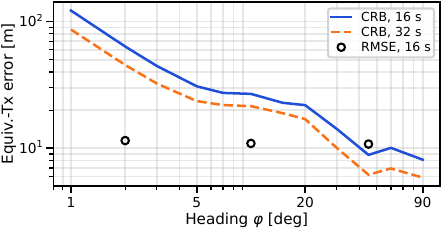}{(c) Degenerate motion}
    \vspace{-1ex}
    \caption{(a) Scalar virtual-Tx CRB \(B_a\) versus window length across deployment geometries; (b) local accuracy bound including the motion prior versus the error scale of the CSI-derived parameters \(\kappa\); (c) the same bound versus the heading angle \(\varphi\) from the radial direction, with Monte Carlo RMSE at \(16\) s.}
    \label{fig:sim_crb}
    \vspace{-1.3em}
\end{figure*}

\begin{figure*}[!t]
    \centering
    \newcommand{\simcell}[2]{%
    \begin{minipage}[t]{0.326\textwidth}
        \centering
        \includegraphics[width=\linewidth]{#1}\\[-1.55ex]
        {\footnotesize #2}
    \end{minipage}}
        \simcell{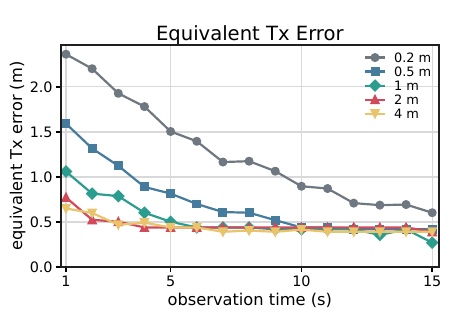}{(a) Virtual Tx error}\hfill
        \simcell{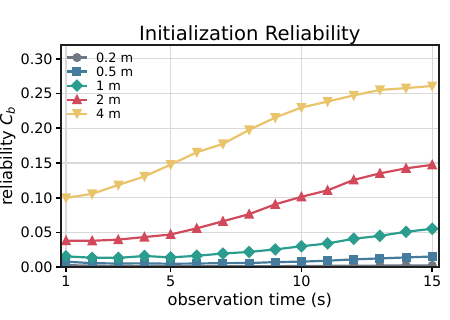}{(b) Initialization confidence}\hfill
        \simcell{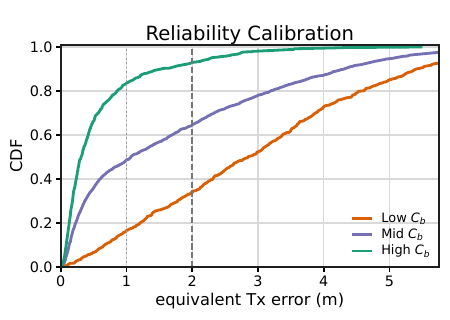}{(c) Reliability calibration}
    \vspace{-1.2ex}
    \caption{Results under varying motion extents, observation times, and randomized deployments. Panels (a) and (b) show median virtual Tx error and initialization confidence \(C_b\); panel (c) shows the virtual Tx error CDF for trials grouped by \(C_b\).}
    \label{fig:sim_motion_reliability}
    \vspace{-1.5em}
\end{figure*}

\section{Simulation Results}
\label{sec:simulations}

\subsection{Parameter Settings}

Table~\ref{tab:experimental_parameters} lists the settings used throughout the CoTrack pipeline. During development, we chose the residual scales to balance the normalized measurement errors and assigned the largest weight to Doppler because, unlike delay and AoA, it contains no unknown reference bias. The virtual-Tx starting points sample several radii and bearings around both the Rx and \(\mathbf c(u)\). The motion bounds reflect the experimental walking area and typical human motion, while the remaining parameters implement the selection and update rules in Section~\ref{sec:cotrack}. All settings are fixed across \mbox{the complete-pipeline simulations and experiments.}

\vspace{-0.8em}
\subsection{Simulation Setup}

\looseness=-1
We first conduct simulations to systematically evaluate the effects of motion extent, observation time, and deployment geometry on CoTrack. The simulations use the coordinate system and initial sensing-parameter model in \eqref{eq:observation_model}, with the receiver at the origin, and generate virtual Tx positions, target trajectories, and virtual Tx candidate searches over the same sensing region used in the real experiments. Given a simulated virtual Tx position \(\mathbf a\), trajectory \(\{\mathbf p_t\}\), and velocity sequence \(\{\mathbf v_t\}\), clean delay-equivalent length, AoA, and Doppler velocity values are generated from the geometric entries of \eqref{eq:predicted_observation}, then corrupted by constant delay/AoA biases and independent zero-mean Gaussian noise at the path-parameter level; the trajectory shape, bias realization, and noise realization are resampled across trials. The simulated stream is fed to the unmodified CoTrack estimator of Section~\ref{sec:cotrack}, and evaluation metrics compare the estimated virtual Tx position or trajectory with the simulated geometry.
The bound validation fixes the angular ambiguity described in Section~\ref{sec:identifiability}. Its baseline standard deviations are \((\tau_d,\tau_\theta,\tau_\nu)=(0.10~{\rm m},2^\circ,0.075~{\rm m/s})\), and \(\kappa\) scales all three. The bound-validation trials use \(\boldsymbol\Lambda=\operatorname{diag}(1/\tau_d,1/\tau_\theta,1/\tau_\nu)\) for both the Monte Carlo estimator and the bound. The acceleration-prior precision is \(\lambda_r/a_0^2\) per component. In Fig.~\ref{fig:sim_crb}(a), a Gaussian prior with a standard deviation of \(1\) mm constrains the virtual-Tx position along the Rx--virtual-Tx direction. Fig.~\ref{fig:sim_crb}(b) and (c) report the local accuracy bound including the motion prior. Position and bias limits constrain only the optimizer. Monte Carlo trials start \eqref{eq:vectorized_ls} close to the true solution.

\vspace{-0.5em}
\subsection{CRB and Identifiability Validation}
\label{sec:simulation_crb}

Fig.~\ref{fig:sim_crb} evaluates three predictions of Section~\ref{sec:identifiability}: the effects of observation geometry and window length, errors in the CSI-derived parameters, and radial motion on virtual-Tx accuracy bounds. In Fig.~\ref{fig:sim_crb}(a), the virtual-Tx CRB varies by over an order of magnitude across deployment geometries and walking patterns: favorable combinations reach sub-meter bounds within about \(12\) s, whereas unfavorable ones require several tens of seconds, which is why CoTrack grows the initialization window instead of committing after a fixed time. In Fig.~\ref{fig:sim_crb}(b), the bounds grow by less than \(2\times\) over a \(16\times\) increase in the parameter-estimation error scale \(\kappa\), since the motion prior limits the degradation, unlike the CRB without the motion prior, which increases linearly with the parameter-estimation error scale (dotted reference); the Monte Carlo RMSE tracks the bound for \(\kappa\ge1\) and is prior- and initialization-limited below. In Fig.~\ref{fig:sim_crb}(c), the local bound including the motion prior inflates by about \(15\times\) as the walking direction approaches the radial direction from the virtual Tx, consistent with the degenerate-motion remark in Section~\ref{sec:identifiability}; the Monte Carlo RMSE tracks the bound where it is informative (\(\varphi=45^{\circ}\)) and saturates at the prior-limited level where the bound exceeds the feasible-region scale. The corresponding candidate-concentration result is validated over \(3000\) randomized geometries in Fig.~\ref{fig:sim_motion_reliability}(c). In weakly identifiable regimes the Monte Carlo RMSE can depart from the bound, as the estimate is shaped by \mbox{the priors and the initialization.}

\makeatletter
\setlength{\@dblfptop}{0pt}
\makeatother
\begin{figure*}[!t]
    \centering
    \newcommand{\setuppanel}[2]{%
    \begin{minipage}[t]{0.247\textwidth}
        \centering
        \includegraphics[width=\linewidth]{#1}\\[-1.55ex]
        {\footnotesize #2}
    \end{minipage}}
    \setuppanel{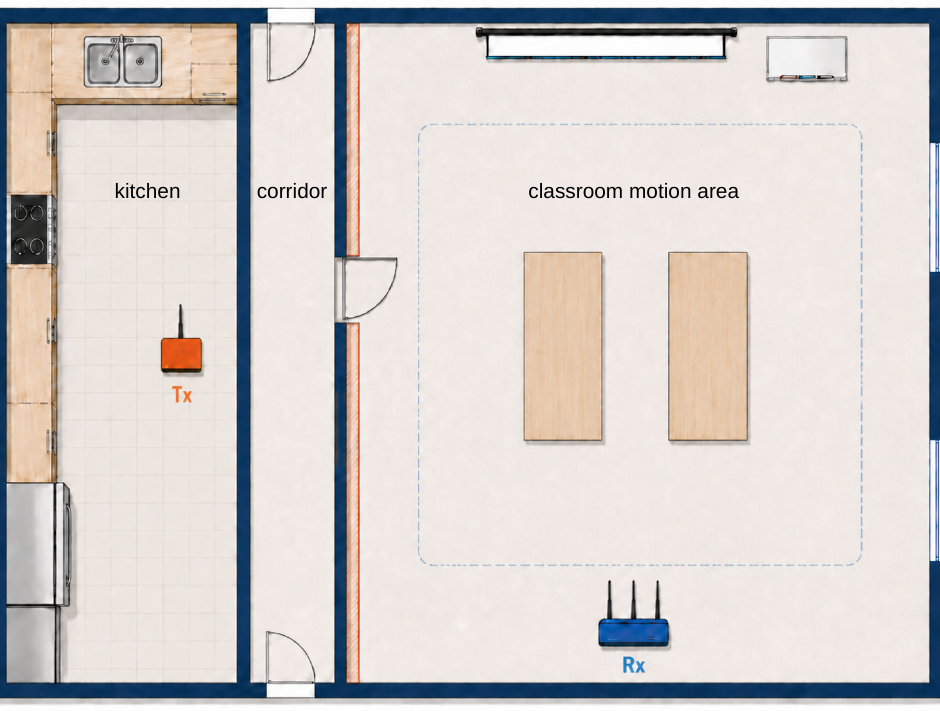}{(a) Floor-plan schematic}%
    \hspace{0.003\textwidth}%
    \setuppanel{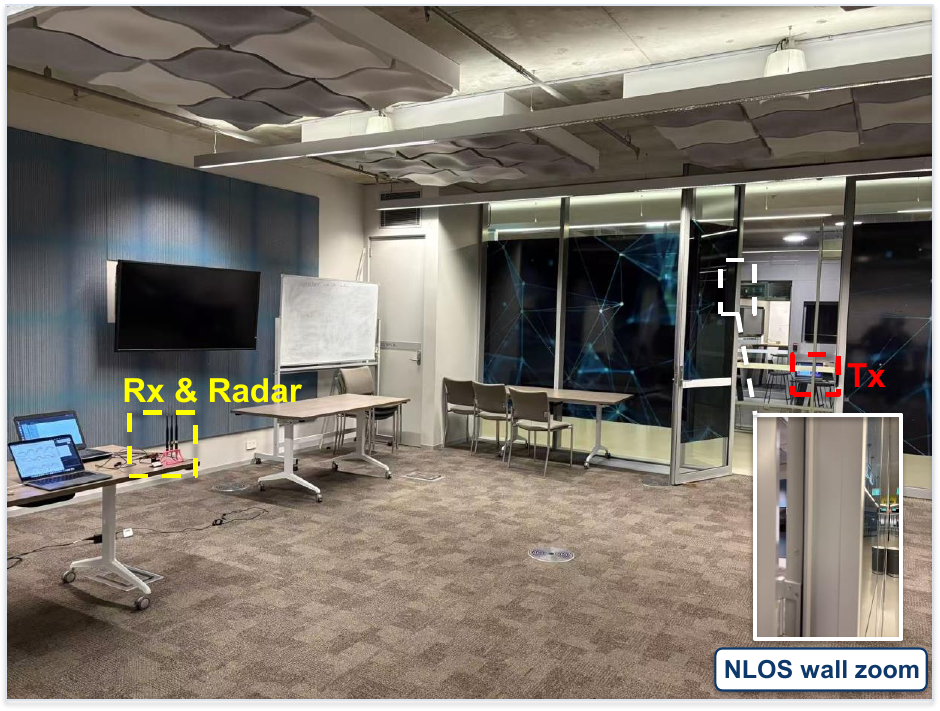}{(b) Cross-room NLOS setup}%
    \hspace{0.003\textwidth}%
    \setuppanel{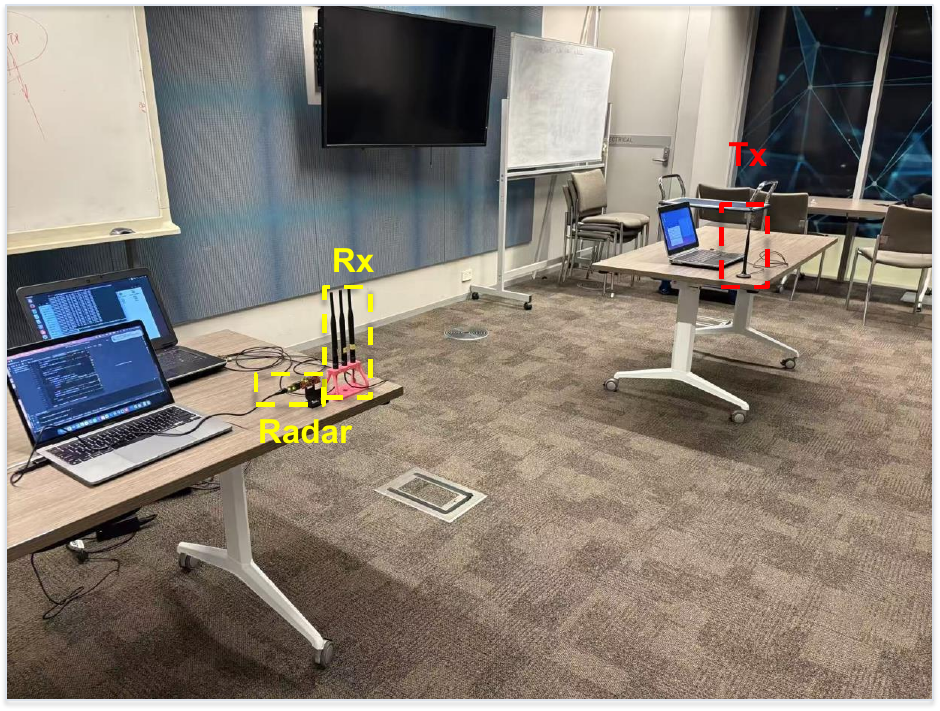}{(c) LOS setup}%
    \hspace{0.003\textwidth}%
    \setuppanel{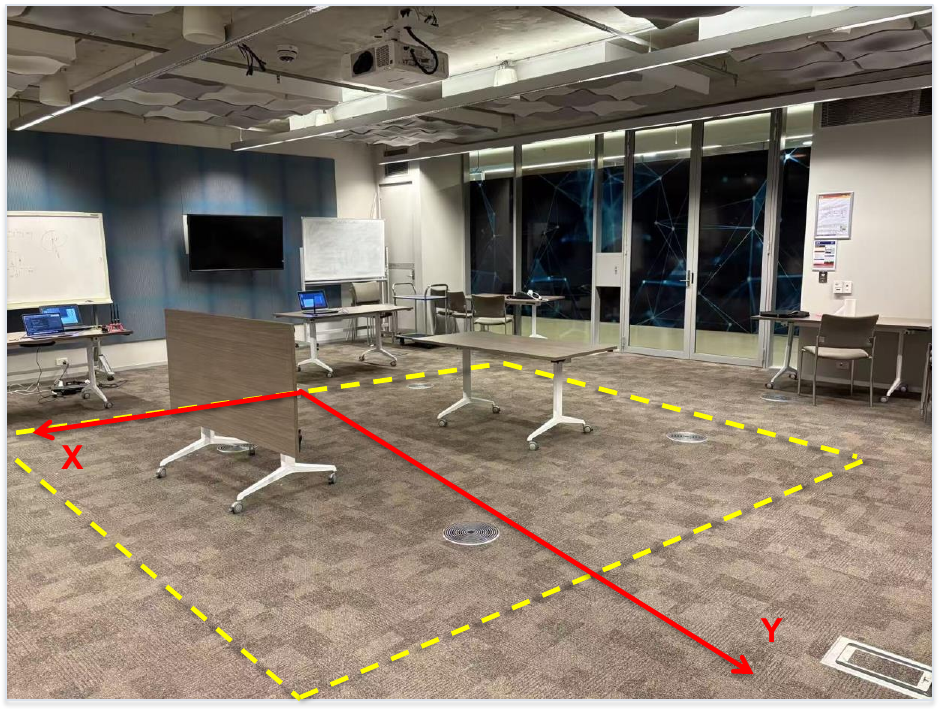}{(d) Motion region}
    \vspace{-1.5em}
    \caption{Experimental setup. The floor-plan schematic shows the kitchen, corridor, classroom, and Tx/Rx placement in the cross-room NLOS setting.}
    
    \label{fig:experimental_setup}
    \vspace{-1.5em}
\end{figure*}

\subsection{Motion Diversity and Initialization Confidence}
Using the simulation setup above, we first evaluate how observation time and motion extent affect the virtual Tx initialization and its confidence score \(C_b\). The simulated virtual Tx position is \((-2.25,0.35)\) m, and the target follows randomized smooth trajectories around the sensing region. For Fig.~\ref{fig:sim_motion_reliability}(a) and (b), the trajectory span varies over \(0.2\), \(0.5\), \(1\), \(2\), and \(4\) m, the initialization CPI window grows from \(1\) s to \(15\) s, and each curve reports the median over \(180\) randomized trials, with delay/AoA biases of \(1.15\) m and \(-7^\circ\) (trial-to-trial jitters \(0.08\) m and \(1.5^\circ\)) and independent zero-mean Gaussian parameter-estimation noise of \(0.10\) m, \(2^\circ\), and \(0.075\) m/s. Fig.~\ref{fig:sim_motion_reliability}(a) shows that the virtual Tx error can decrease even for compact motion as parameter estimates accumulate, since repeated measurements reduce random noise. However, \(C_b\) in Fig.~\ref{fig:sim_motion_reliability}(b) remains low for the \(0.2\) m and \(0.5\) m spans, reflecting limited spatial concentration or score contrast among the low-score candidates, whereas, with the feasible region fixed, the \(2\) m and \(4\) m spans yield both lower error and a clearly increasing confidence score because larger motion provides stronger delay, AoA, and Doppler variation. The resulting \(C_b\) indicates reliable initialization.

\vspace{-0.7em}
\subsection{Randomized-Geometry Reliability Calibration}

\looseness=-1
We further test whether \(C_b\) predicts output reliability when the deployment geometry changes. Fig.~\ref{fig:sim_motion_reliability}(c) uses \(3000\) randomized-geometry trials with randomized virtual Tx distance and bearing, bias/noise realizations, motion spans of \(1\), \(2\), \(4\), and \(6\) m, and observation times of \(5\), \(8\), \(12\), and \(15\) s. The trials are grouped only by the data-derived \(C_b\) value, while the ground-truth virtual Tx position is used only for evaluation. The high-\(C_b\) group achieves a median virtual Tx error of \(0.31\) m, with \(83.5\%\) and \(92.9\%\) of trials within \(1\) m and \(2\) m, respectively; the middle-\(C_b\) group has a median error of \(1.07\) m, while the low-\(C_b\) group has a median error of \(2.85\) m. This separation indicates that \(C_b\) is not merely a numerical score attached to the selected candidate, but a reliability indicator for deciding whether the current virtual Tx estimate should be output: high \(C_b\) corresponds to a stable and accurate estimate, whereas low \(C_b\) reflects insufficient motion diversity or weak geometric excitation and should prevent committing the virtual Tx estimate.

\section{Real-World Experiments}
\label{sec:experiments}

\subsection{Platform and Scenes}
CoTrack is evaluated with one WiFi transmitter-receiver link. The WiFi devices use Intel 5300 network interface cards~\cite{halperin2010predictable,halperin2014linux80211ncsitool} at \(5.32\) GHz with \(20\)-MHz bandwidth, and CSI is sampled at \(1\) kHz. The transmitter uses one antenna, while the receiver uses a three-antenna uniform linear array (ULA) with half-wavelength spacing. An independent TI millimeter-wave radar is placed close to the WiFi receiver and oriented toward the walking area. The radar operates at \(60\) GHz with a \(33.3\)-ms frame period, \(4.7\)-cm range resolution, about \(10\)~m maximum unambiguous range, and \(2.24\) m/s maximum radial velocity. It provides Cartesian detections and radial velocity only for reference construction, not as CoTrack input.

Fig.~\ref{fig:experimental_setup}(a) shows the overall experimental layout. The classroom spans approximately \(12\) m laterally, about \(6\) m to each side of the receiver, and approximately \(12\) m in the forward direction. A corridor of about \(2\) m width separates the classroom from a kitchen-side transmitter area. In the cross-room NLOS setting, shown in Fig.~\ref{fig:experimental_setup}(b), the transmitter is placed in the kitchen-side area behind the intervening wall, approximately \(8.7\) m to the receiver's left and about \(9\) m from the receiver. In the LOS setting, shown in Fig.~\ref{fig:experimental_setup}(c), the transmitter is placed in the same classroom, about \(2.5\) m from the receiver and at a bearing of about \(82^\circ\) to the receiver's left. In both settings, a single walking target moves only within the classroom sensing region illustrated in Fig.~\ref{fig:experimental_setup}(d).

\begin{figure*}[!t]
    \centering
    \newcommand{\resultcell}[2]{%
        \begin{minipage}[t]{\linewidth}
            \centering
            \includegraphics[width=\linewidth]{#1}\\[-1.35ex]
            {\footnotesize #2}
        \end{minipage}}
    \setlength{\tabcolsep}{0pt}
    \begin{tabular}{@{}p{0.328\textwidth}@{\hspace{0.004\textwidth}}p{0.328\textwidth}@{\hspace{0.004\textwidth}}p{0.328\textwidth}@{}}
        \resultcell{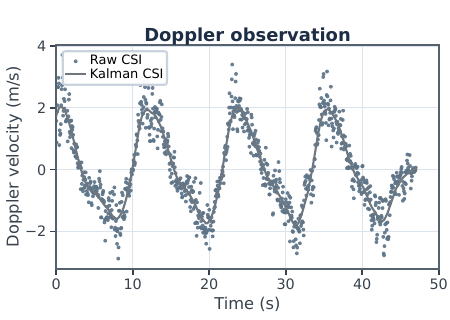}{(a) LOS Doppler} &
        \resultcell{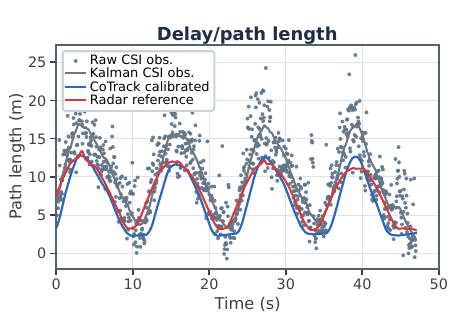}{(b) LOS delay} &
        \resultcell{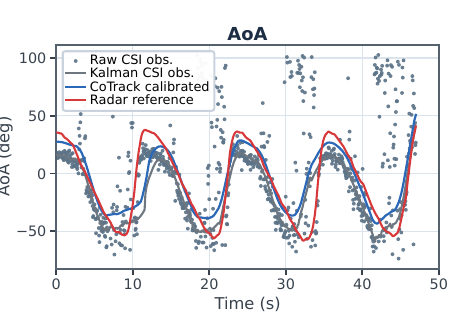}{(c) LOS AoA}
        \\[-0.05ex]
        \resultcell{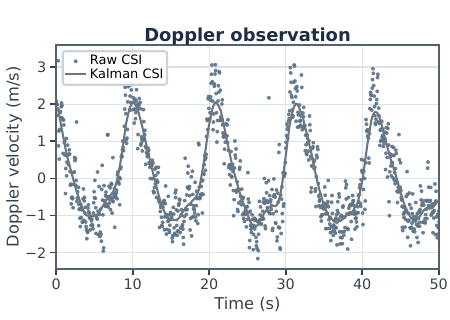}{(d) NLOS Doppler} &
        \resultcell{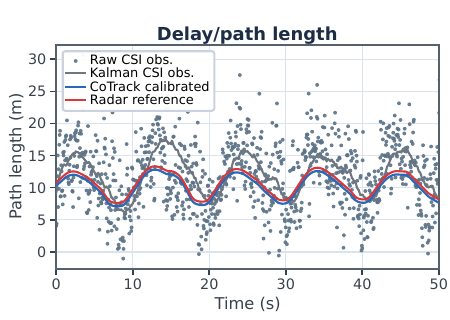}{(e) NLOS delay} &
        \resultcell{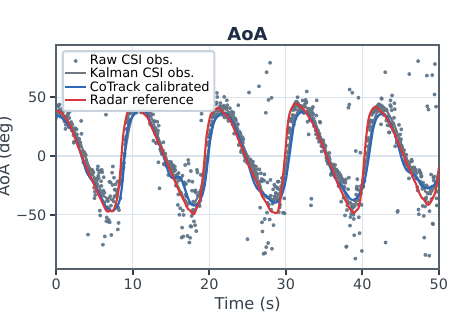}{(f) NLOS AoA}
        \\[-0.05ex]
        \resultcell{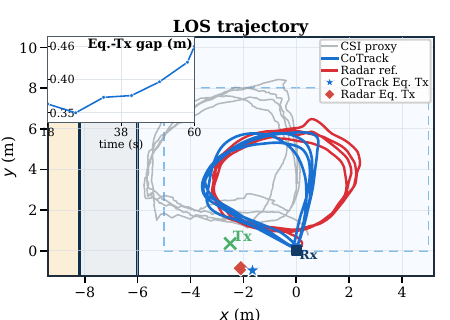}{(g) Representative LOS} &
        \resultcell{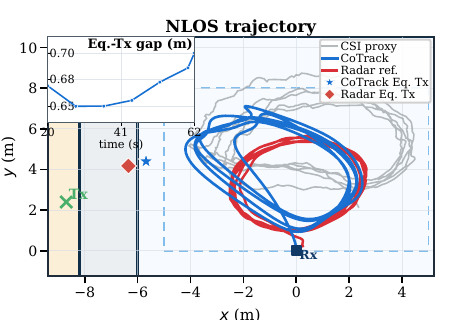}{(h) Cross-room NLOS} &
        \resultcell{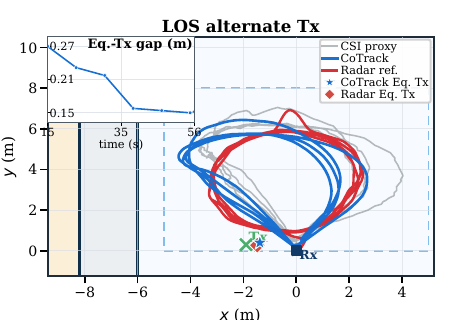}{(i) LOS, alternate Tx}
        \\[-0.05ex]
        \resultcell{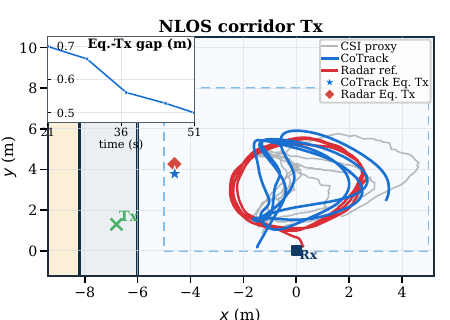}{(j) Corridor-side NLOS Tx} &
        \resultcell{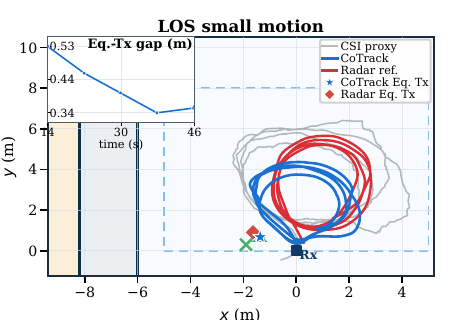}{(k) LOS small motion} &
        \resultcell{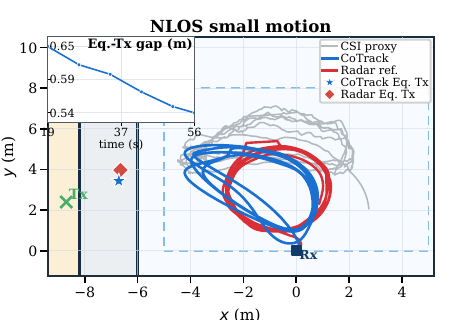}{(l) NLOS small motion}
    \end{tabular}
    \vspace{-0.8ex}
    \caption{Experimental results. Panels (a)--(f) show Doppler, delay, and AoA; panels (g)--(l) show the trajectories and virtual-Tx positions, with insets showing the virtual-Tx discrepancy over time. In (a)--(f), blue, gray, and red denote CoTrack-calibrated results, CSI estimates, and radar reference; in (g)--(l), they denote CoTrack, CSI range proxy, and radar reference, with blue stars/red diamonds marking CoTrack/radar-aided virtual Tx positions.}
    \label{fig:results_cases}
    \vspace{-1.5em}
\end{figure*}

\vspace{-1em}
\subsection{Evaluation References and Initial Estimates}
Short-range millimeter-wave radars have been widely used for indoor human tracking and point-cloud-based trajectory estimation~\cite{pearce2023multi}. The radar measurements provide accurate absolute \(x\)-\(y\) observations of the moving person for evaluation. Raw point detections are grouped by frame, and candidate human clusters are generated with a fixed-radius neighborhood. The selected cluster balances temporal continuity, point density, and signal-to-noise ratio (SNR), and its SNR-weighted center is used as the frame-level target position. We then apply local Hampel outlier rejection, kinematic repair for isolated jumps, interpolation of repaired samples, triangular moving smoothing, and local time-window regression to obtain the reference position and velocity. The radar trajectory is rigidly transformed to the Rx frame and interpolated to WiFi times.

The measured physical Tx coordinate is shown in LOS scenes only as a physical reference. Quantitative trajectory errors use the radar trajectory, whereas virtual-Tx discrepancies use the radar-aided reference; physical-Tx discrepancies reported in LOS are supplementary checks. In NLOS scenes, the physical Tx coordinate is not an appropriate virtual-Tx ground-truth target, because the virtual Tx position may represent the dominant transmitter-side propagation path rather than the physical antenna location. We obtain the radar-aided reference \(\mathbf a^{\rm R}\) by fixing the trajectory in \eqref{eq:joint_problem} to the radar trajectory and fitting \((\mathbf a,\mathbf b)\). This removes trajectory uncertainty and yields the virtual Tx supported by the CSI observations when the target trajectory is known. During estimation, the approximate room layout bounds the target motion area, while the virtual Tx uses the broader numerical bounds in Table~\ref{tab:experimental_parameters}.

\vspace{-1em}
\subsection{LOS and Cross-Room NLOS Results}
\looseness=-1
We first compare representative LOS and cross-room NLOS results with the radar reference. Fig.~\ref{fig:results_cases}(a)--(f) shows the estimated sensing parameters. A Kalman filter suppresses frame-level outliers in the raw CSI-derived delay, AoA, and Doppler estimates. Because Doppler is already an absolute velocity measurement, its panels show both the raw and filtered CSI estimates. Delay and AoA, by contrast, remain expressed relative to their unknown reference path before CoTrack calibration. Their direct \(x\)-\(y\) mapping is therefore shown only as a CSI range proxy, whereas the CoTrack curves are plotted after removal of the estimated delay and AoA biases. Under both LOS and NLOS propagation, the calibrated CoTrack estimates follow the radar-derived geometric trends more closely than the biased CSI measurements.

\looseness=-1
Fig.~\ref{fig:results_cases}(g) and Fig.~\ref{fig:results_cases}(h) compare the recovered trajectories and virtual Tx positions. The trajectory error is \(e_t=\|\widehat{\mathbf p}_t-\mathbf p_t^{\rm R}\|_2\) against the time-synchronized radar reference \(\mathbf p_t^{\rm R}\). The virtual-Tx discrepancy \(\|\widehat{\mathbf a}-\mathbf a^{\rm R}\|_2\) evaluates whether the joint estimator recovers the same propagation reference while estimating the trajectory simultaneously. Both metrics are computed in the common Rx-centered frame over the synchronized samples shown. The directly returned feasible solution is used without radar-based gauge alignment. In the LOS experiment, the median trajectory error, root-mean-square error (RMSE), and \(80\)th-percentile error are \(1.61\) m, \(1.65\) m, and \(2.04\) m, with a virtual-Tx discrepancy of \(0.46\) m; in the cross-room NLOS experiment, they are \(0.95\) m, \(1.80\) m, and \(2.61\) m, with a \(0.70\) m discrepancy. Over the two runs, the discrepancies vary by \(0.10\) m and \(0.05\) m in (g) and (h), respectively. They may rise or fall because online updates minimize CSI fitting error rather than the plotted discrepancy; their small variation shows that the virtual-Tx estimates remain stable.

\vspace{-1em}
\subsection{Effect of Transmitter Placement}
\looseness=-1
We next change the transmitter-side geometry while retaining a sufficiently large walking trajectory to produce appreciable delay, AoA, and Doppler variation. This experiment tests whether CoTrack is specific to one Tx placement. In the alternate LOS case of Fig.~\ref{fig:results_cases}(i), CoTrack estimates the physical Tx with an error of \(0.53\) m and differs from the radar-aided virtual-Tx reference by \(0.15\) m. The median trajectory error and RMSE are \(1.42\) m and \(1.71\) m, respectively. In the corridor-side NLOS case of Fig.~\ref{fig:results_cases}(j), these trajectory errors decrease to \(0.88\) m and \(1.09\) m, while the virtual-Tx discrepancy is \(0.48\) m.

\vspace{-1em}
\subsection{Effect of Motion Extent}
\looseness=-1
We further evaluate smaller walking trajectories, which weaken the motion-induced geometric diversity that CoTrack relies on. Fig.~\ref{fig:results_cases}(k) and Fig.~\ref{fig:results_cases}(l) show the LOS and NLOS results. In the small-motion LOS experiment, CoTrack estimates the physical transmitter with \(0.66\) m error, with a trajectory median error and RMSE of \(1.34\) m and \(1.43\) m. Its radar-aided discrepancy is \(0.36\) m. In the small-motion NLOS experiment, the trajectory median error and RMSE are \(0.93\) m and \(1.37\) m, with a \(0.54\) m virtual-Tx discrepancy. Taken together, the median/\(80\)th-percentile values are \(1.14/1.42\) m for the six trajectory median errors and \(0.47/0.54\) m for the six final virtual-Tx discrepancies. The online virtual-Tx estimates change by only \(0.11\)--\(0.37\) m from first to last. In all three pairs, NLOS has lower trajectory errors but larger virtual-Tx discrepancies. The cases use different Tx placements and walking paths; these differences can make tracking easier while reducing the Tx-to-person direction changes used to estimate the virtual Tx.

\vspace{-1em}
\section{Discussion and Conclusion}
\label{sec:conclusion}

This paper has presented CoTrack, a passive WiFi tracking scheme for LOS and cross-room NLOS deployments in which the Tx position is unknown. Using only coarse feasible-region bounds, CoTrack jointly estimates a virtual Tx, the delay and AoA biases, and the human trajectory from CSI-derived delay, AoA, and Doppler measurements. The analysis characterizes the conditions for local identifiability, while the experiments demonstrate robust operation across the tested \mbox{motion patterns and transmitter placements.}

In NLOS, the virtual Tx represents the effective propagation geometry created by penetration, reflection, and diffraction; it need not represent the physical antenna position. Performance depends on both motion diversity and the accuracy of the CSI-derived parameters. Limited motion provides insufficient geometric diversity to distinguish competing virtual-Tx hypotheses, whereas persistent parameter biases propagate into the recovered trajectory. Future work will improve the front-end parameter estimates and extend CoTrack to multiple targets by exploiting a virtual Tx and reference biases \mbox{shared across separable human-induced paths.}

\vspace{-1em}
\appendix[Proofs and Derivations]
\label{app:interface}

\subsection{Proof of Proposition~\ref{prop:existence}: Local Virtual-Tx Approximation}
\label{app:virtual_tx_proof}

{
The unit-gradient condition is the local eikonal relation for a propagation-length field \cite{balanis2012advanced}. For a single planar specular reflection, for example, the mirror image of the physical Tx is an exact virtual source \cite{meissner2010uwb}.

Let \(\boldsymbol\delta=\mathbf p-\mathbf p_0\), \(\mathbf u=\nabla L(\mathbf p_0)\), and \(\mathbf K=\nabla^2L(\mathbf p_0)\). Because \(L\) is twice continuously differentiable,
\begin{equation}
L(\mathbf p)
=L(\mathbf p_0)+\mathbf u^T\boldsymbol\delta
+\frac{1}{2}\boldsymbol\delta^T\mathbf K\boldsymbol\delta
+o(\|\boldsymbol\delta\|_2^2).
\label{eq:physical_path_expansion}
\end{equation}
The unit-gradient assumption gives \(\|\mathbf u\|_2=1\). Differentiating \(\|\nabla L(\mathbf p)\|_2^2=1\) at \(\mathbf p_0\) also gives \(\mathbf K\mathbf u=\mathbf0\).

For any \(r>0\), set \(\mathbf a=\mathbf p_0-r\mathbf u\). Then \(\mathbf p-\mathbf a=r\mathbf u+\boldsymbol\delta\), and expansion of its norm around \(\boldsymbol\delta=\mathbf0\) yields
\begin{equation}
\begin{aligned}
\|\mathbf p-\mathbf a\|_2
=r+\mathbf u^T\boldsymbol\delta
+\frac{1}{2r}\boldsymbol\delta^T
(\mathbf I-\mathbf u\mathbf u^T)\boldsymbol\delta
+o(\|\boldsymbol\delta\|_2^2).
\end{aligned}
\label{eq:virtual_path_expansion}
\end{equation}
Choosing \(C=L(\mathbf p_0)-r\) makes the constant terms in \eqref{eq:physical_path_expansion} and \eqref{eq:virtual_path_expansion} equal. Their linear terms are already identical, which proves \eqref{eq:virtual_tx_first_order_match} and \eqref{eq:virtual_tx_taylor_match}. If \(\mathbf K=(\mathbf I-\mathbf u\mathbf u^T)/r\), their quadratic terms also agree, leaving only \(o(\|\boldsymbol\delta\|_2^2)\).

Finally, consider motion \(\mathbf p(t_{\rm c})\) with velocity \(\mathbf v=\mathrm d\mathbf p(t_{\rm c})/\mathrm d t_{\rm c}\). At \(\mathbf p_0\), both transmitter-side models have gradient \(\mathbf u\), so both have instantaneous path-length rate \(\mathbf u^T\mathbf v\) and Doppler contribution \(-\mathbf u^T\mathbf v/\lambda\). Because the person-to-Rx segment is identical in the two models, their total Doppler values also agree at \(\mathbf p_0\).
}

\vspace{-0.5em}
\subsection{Jacobian Blocks of the FIM}
\label{app:fim_jacobians}

{
Let \(r_{R,t}=\|\mathbf p_t\|_2\), \(r_{A,t}=\|\mathbf p_t-\mathbf a\|_2\), \(\mathbf P_{R,t}=\mathbf I-\mathbf u_{R,t}\mathbf u_{R,t}^T\), and \(\mathbf P_{A,t}=\mathbf I-\mathbf u_{A,t}\mathbf u_{A,t}^T\). Using \(\partial\mathbf u_{R,t}/\partial\mathbf p_t=\mathbf P_{R,t}/r_{R,t}\) and \(\partial\mathbf u_{A,t}/\partial\mathbf p_t=\mathbf P_{A,t}/r_{A,t}=-\partial\mathbf u_{A,t}/\partial\mathbf a\), the nonzero entries of \(\mathbf J_t\) follow from \eqref{eq:predicted_observation}. For the delay row, \(\partial[\mathbf h_t]_1/\partial\mathbf a=-\mathbf u_{A,t}^T-\mathbf a^T/\|\mathbf a\|_2\), \(\partial[\mathbf h_t]_1/\partial\mathbf p_t=(\mathbf u_{R,t}+\mathbf u_{A,t})^T\), and \(\partial[\mathbf h_t]_1/\partial b_d=1\). For the AoA row, \(\partial[\mathbf h_t]_2/\partial\mathbf p_t=[y_t,\,-x_t]/r_{R,t}^{2}\) and \(\partial[\mathbf h_t]_2/\partial b_\theta=1\). For an interior CPI, the Doppler row has \(\partial[\mathbf h_t]_3/\partial\mathbf a=-\mathbf v_t^{T}\mathbf P_{A,t}/r_{A,t}\), \(\partial[\mathbf h_t]_3/\partial\mathbf p_t=\mathbf v_t^{T}(\mathbf P_{R,t}/r_{R,t}+\mathbf P_{A,t}/r_{A,t})\), and \(\partial[\mathbf h_t]_3/\partial\mathbf p_{t\pm1}=\pm(\mathbf u_{R,t}+\mathbf u_{A,t})^T/(2\Delta T)\). Boundary entries use the one-sided rows of \(\mathbf D_1\); all other entries are zero.
}

\vspace{-0.5em}
\subsection{Proof of the Degenerate-Motion Remark}
\label{app:degenerate_motion}

Let \(\mathbf u_{A,t}\equiv\mathbf u\) for all \(t\in\mathcal W\), \(\mathbf a'=\mathbf a+\epsilon\mathbf u\), and \(b_d'=b_d+\epsilon+\|\mathbf a'\|_2-\|\mathbf a\|_2\), where \(0<\epsilon<\min_t r_{A,t}\). Since \(\mathbf p_t-\mathbf a=r_{A,t}\mathbf u\), \(\mathbf p_t-\mathbf a'=(r_{A,t}-\epsilon)\mathbf u\), and the delay prediction satisfies \(r_{R,t}+(r_{A,t}-\epsilon)-\|\mathbf a'\|_2+b_d'=r_{R,t}+r_{A,t}-\|\mathbf a\|_2+b_d\). The AoA is independent of \(\mathbf a\), and the unchanged \(\mathbf u_{A,t}\) preserves the Doppler prediction. Hence, both parameter sets give identical predictions, making the likelihood constant and \(\mathbf F\) singular.

\vspace{-0.5em}
\subsection{Proof of Theorem~\ref{thm:curvature}}
\label{app:profiled_curvature}

Fix \(b_\theta=0\), let \(\mathbf y=[b_d,\operatorname{vec}(\mathbf P_{\mathcal W})^T]^T\), and define \(F_{\mathcal W}(\mathbf a)=\min_{\mathbf y}\|\mathbf r_{\mathcal W}(\mathbf a,\mathbf y)\|_2^2\), where \(\mathbf r_{\mathcal W}\) is the residual under this constraint and minimization is restricted to a neighborhood of the fitted \(\mathbf y\). At an interior differentiable fit, let \(\mathbf J_a\) and \(\mathbf J_y\) be the corresponding Jacobians, evaluated at that fit, and \(\boldsymbol\Pi_y^\perp=\mathbf I-\mathbf J_y(\mathbf J_y^T\mathbf J_y)^\dagger\mathbf J_y^T\), where \((\cdot)^\dagger\) is the Moore--Penrose pseudoinverse. For small residuals, the profiled Gauss--Newton curvature \(\mathbf H_a=\mathbf J_a^T\boldsymbol\Pi_y^\perp\mathbf J_a\) satisfies \(\nabla^2F_{\mathcal W}(\mathbf a)\approx2\mathbf H_a\).

For a small perturbation \(\delta\mathbf a\), least squares removes the component of \(\mathbf J_a\delta\mathbf a\) in \(\operatorname{range}(\mathbf J_y)\), giving
\begin{equation}
    \min_{\delta\mathbf y}
    \left\|
    \mathbf J_a\delta\mathbf a
    +
    \mathbf J_y\delta\mathbf y
    \right\|_2^2
    =
    \left\|
    \boldsymbol\Pi_y^\perp
    \mathbf J_a\delta\mathbf a
    \right\|_2^2
    =
    \delta\mathbf a^T\mathbf H_a\delta\mathbf a.
    \label{eq:profiled_curvature}
\end{equation}

Since \(\mathbf H_a\) is positive semidefinite, this minimum is zero exactly when \(\mathbf H_a\delta\mathbf a=\mathbf0\), proving part~(i).

Under the assumptions of part~(ii), \(\nabla F_{\mathcal W}(\mathbf a^\star)=\mathbf0\), so Taylor expansion and the Hessian bound give \(F_{\mathcal W}(\mathbf a_i)-F_{\mathcal W}(\mathbf a^\star)\ge\lambda_a\|\mathbf a_i-\mathbf a^\star\|_2^2/2+o(\|\mathbf a_i-\mathbf a^\star\|_2^2)\). Combining this with \(F_{\mathcal W}(\mathbf a_i)-F_{\mathcal W}(\mathbf a^\star)\le\Delta_J\) for nearby candidates yields
\begin{equation}
    \|\mathbf a_i-\mathbf a^\star\|_2
    \le
    \sqrt{\frac{2\Delta_J}{\lambda_a}}
    +
    o(\sqrt{\Delta_J}),
    \label{eq:basin_bound}
\end{equation}
where \(o(\sqrt{\Delta_J})\) denotes higher-order terms as \(\Delta_J\to0\).

When \(\boldsymbol\Lambda^T\boldsymbol\Lambda=\boldsymbol\Sigma_e^{-1}\) and \(\lambda_r=0\), block elimination gives \(\mathbf H_a=\mathbf F_{aa}-\mathbf F_{ay}\mathbf F_{yy}^{\dagger}\mathbf F_{ya}\); its inverse on the identifiable subspace after fixing the angular ambiguity is the CRB.

\makeatletter
\newcommand{\BibLooseness}[2]{\expandafter\def\csname CT@bibloose@#1\endcsname{#2}}
\newcommand{\CT@bibfit}[1]{\looseness=0\relax\rightskip=0pt\relax\ifcsname CT@bibloose@#1\endcsname\looseness=\csname CT@bibloose@#1\endcsname\relax\fi\ifcsname CT@bibpatch@#1\endcsname\rightskip=0pt plus 1em\relax\fi}
\newcommand{\BibNoBreak}[3]{\expandafter\def\csname CT@bibpatch@#1\endcsname{\patchcmd{\CT@body}{#2}{#3}{}{\PackageWarning{bibfit}{No line-break adjustment applied to #1; bibliography text may have changed}}}}
\newcommand{\BIBdecl}{\let\CT@origbibitem\@bibitem\long\def\@bibitem##1##2\par{\CT@origbibitem{##1}\CT@bibfit{##1}\def\CT@body{##2}\ifcsname CT@bibpatch@##1\endcsname\csname CT@bibpatch@##1\endcsname\fi\CT@body\par}}
\makeatother
\BibNoBreak{shen2024multiroom}{vol.~73, pp. 1--17, 2024.}{\mbox{vol.~73, pp. 1--17, 2024.}}
\BibNoBreak{xie2019mdtrack}{\emph{The 25th Annual International Conference on Mobile Computing and Networking}, 2019, pp. 1--16.}{\emph{The 25th Annual International Conference on Mobile Computing} \mbox{\emph{and Networking}, 2019, pp. 1--16.}}
\BibNoBreak{wang2026wifi}{\emph{IEEE Transactions on Mobile Computing}, 2026.}{\emph{IEEE} \mbox{\emph{Transactions on Mobile Computing}, 2026.}}
\BibNoBreak{xu2024radio}{position,'' \emph{arXiv preprint arXiv:2411.04398}, 2024.}{\mbox{position,'' \emph{arXiv preprint arXiv:2411.04398}, 2024.}}
\BibNoBreak{chen2023device}{no.~11, pp. 1327--1340, 2023.}{\mbox{no.~11, pp. 1327--1340, 2023.}}
\BibNoBreak{zhang2020calwifi}{no.~1, pp. 661--664, 2019.}{\mbox{no.~1, pp. 661--664, 2019.}}
\BibNoBreak{zubow2021phase}{\emph{2021 16th Annual Conference on Wireless On-demand Network Systems and Services Conference (WONS)}.\hskip 1em plus 0.5em minus 0.4em\relax IEEE, 2021, pp. 1--4.}{\emph{2021 16th Annual Conference on Wireless On-demand Network Systems and Services Conference} \mbox{\emph{(WONS)}.\hskip 1em plus 0.5em minus 0.4em\relax IEEE, 2021, pp. 1--4.}}
\BibNoBreak{meissner2010uwb}{\emph{2010 7th Workshop on Positioning, Navigation and Communication (WPNC)}.\hskip 1em plus 0.5em minus 0.4em\relax IEEE, 2010, pp. 150--156.}{\emph{2010 7th Workshop on Positioning, Navigation and Communication} \mbox{\emph{(WPNC)}.\hskip 1em plus 0.5em minus 0.4em\relax IEEE, 2010, pp. 150--156.}}
\BibLooseness{ma2019wifiCSI}{-1}
\BibLooseness{gu2025csipose}{-1}
\BibLooseness{lee2025wi}{-1}
\BibLooseness{wu2024sensing}{-1}
\BibLooseness{leitinger2019belief}{-1}
\BibLooseness{suraweera2020environment}{-1}
\BibLooseness{kato2025multi}{-1}

\bibliographystyle{IEEEtran}
\bibliography{references}

\end{document}